\documentclass[fleqn,usenatbib]{mnras}
\usepackage{newtxtext,newtxmath}
\usepackage[T1]{fontenc}

\DeclareRobustCommand{\VAN}[3]{#2}
\let\VANthebibliography\thebibliography
\def\thebibliography{\DeclareRobustCommand{\VAN}[3]{##3}\VANthebibliography}

\definecolor{changegreen}{RGB}{0,110,0}
\usepackage[normalem]{ulem}

\usepackage{amsmath}
\usepackage{physics}
\usepackage{graphicx}
\graphicspath{ {./figures/} }
\usepackage[table]{xcolor}
\usepackage{colortbl}
\usepackage{url}
\usepackage{longtable}
\usepackage{multirow}
\usepackage{anyfontsize}

\title[13 Year short GRB catalog]{Expanding the Population of Short Gamma-Ray Transients with a Coherent Fermi/GBM Search. A 13-year catalog of short GRBs}

\author[A. Perera, B. Zackay and T. Venumadhav]{
Ariel Perera,$^{1}$\thanks{E-mail: ariel.perera@weizmann.ac.il}
Barak Zackay$^{1}$
and Tejaswi Venumadhav$^{2,3}$
\\
$^{1}$Department of Particle Physics \& Astrophysics, Weizmann Institute of Science, Rehovot 76100, Israel\\
$^{2}$Department of Physics, University of California at Santa Barbara, Santa Barbara, CA 93106, USA\\
$^{3}$International Centre for Theoretical Sciences, Tata Institute of Fundamental Research, Bangalore 560089, India
}

\date{Accepted 2026 August 20. Received 2026 August 20; in original form 2026 May 29}

\pubyear{2026}

\begin{document}
\label{firstpage}
\pagerange{\pageref{firstpage}--\pageref{lastpage}}
\maketitle

\begin{abstract}
    In this paper, we present an archival search for short gamma-ray bursts (sGRBs) over 13 years (2013-2025) of Fermi/GBM data using a Poisson matched-filter pipeline that performs a fully coherent analysis across all detectors and energy channels, significantly improving sensitivity relative to the onboard triggering algorithms. A central component of the analysis is the empirical estimation of trigger significance using 'timeslided' data, allowing each candidate to be assigned a probability of astrophysical origin. We also developed a new parameter-estimation framework based on the Poisson matched filter, which uses the global structure of the detected event across spectral, temporal, and spatial parameter spaces. This enables us to systematically classify bursts and distinguish between GRBs, soft gamma repeaters, terrestrial gamma-ray flashes, and solar flares. We identify 549 new GRB candidates with high confidence and thousands of magnetar bursts, significantly expanding the known short transient population in GBM data. To further strengthen the significance of the GRB candidates, we performed a targeted follow-up search in Swift/BAT rate data. Applying the followup to all of our triggers - including triggers below the detection threshold yielded 1736 temporally coincident events with association probability above $90\%$. The resulting probabilistically ranked catalog substantially expands the population of short transients detected in GBM data and provides a statistically robust framework for multimessenger searches.
\end{abstract}




\section{Introduction}
    Gamma-ray bursts (GRBs) have been studied for more than five decades, with successive generations of gamma-ray instruments shaping our understanding of these extreme transient phenomena. Early observations established the existence of GRBs (\cite{klebesadel_observations_1973}), while later measurements demonstrated their cosmological origin (\cite{paczynski_gamma-ray_1986, meegan_spatial_1992}) and revealed a bimodal distribution in duration and spectral hardness, enabled by the large sample collected by BATSE (\cite{kouveliotou_identification_1993, kaneko_complete_2006}). The advent of rapid localization and follow-up capabilities with \textit{Swift} (\cite{gehrels_swift_2004}) marked a further milestone, enabling arcminute localizations and systematic afterglow studies.
    
    Among current instruments, the Gamma-ray Burst Monitor (GBM) (\cite{meegan_fermi_2009}) on board the \textit{Fermi} satellite has played a central role in expanding the GRB sample, detecting $\sim240$ GRBs per year, of which $\sim40$ are short-duration bursts (\cite{von_kienlin_fourth_2020}). Most notably, the detection of GRB170817A in temporal coincidence with the gravitational-wave event GW170817 (\cite{goldstein_ordinary_2017, savchenko_integral_2017, abbott_gravitational_2017, abbott_gw170817_2017}) marked the beginning of the multimessenger era in GRB astronomy. 
    
    The first joint detection, GRB-GW 170817, remains the only multimessenger event to date. 
    Despite GBM's success in increasing the GRB sample size, its onboard triggering algorithms are necessarily constrained and are not designed to be optimal for the detection of faint, short-duration bursts. 
    Triggers are generated when the count rates in two or more detectors, for some choices of timescales, exceed fixed thresholds above the background (\cite{von_kienlin_fourth_2020}). 
    This triggering logic does not make full use of the available spectral and directional information, which results in an effective reduction of the signal-to-noise ratio (SNR) by approximately a factor of two compared to a coherent analysis of all detectors (\cite{blackburn_high-energy_2015}). 
    Several approaches have been developed to extend the GRB detection sensitivity of GBM beyond the onboard triggering algorithms. These include targeted searches, which apply methods that exploit coherence, and also use templates for the temporal structure but only within externally motivated time windows \citep{kocevski_analysis_2018}, as well as blind subthreshold searches that apply similar techniques to the continuous data stream without predefined triggers.\footnote{\url{https://gcn.gsfc.nasa.gov/gcn/fermi_gbm_subthresh_archive.html}} Additional methods based on light-curve variability and rate excesses have also been applied \citep{kaneko_11_2026}. While these approaches have demonstrated greater sensitivity than onboard triggering, there remains room for improvement. In particular, the employed template sets provide limited coverage of the full spectral and directional parameter space, the lightcurve analyses often rely on combined count rates rather than fully coherent detector-by-detector information, and the statistical interpretation of low-significance candidates is not derived from empirically measured background distributions. 
    An additional complication is that the GBM data contain a variety of high-energy transient phenomena apart from GRBs. These include soft gamma repeaters (SGRs), terrestrial gamma-ray flashes (TGFs), and solar flares (SFs), many of which can produce short, impulsive signals that overlap in duration and energy with GRBs. In particular, SGR bursts constitute a significant population of short-duration transients in the GBM data and can dominate the trigger rate in some regions of parameter space. As a result, searches must contend with a heterogeneous mixture of transients, motivating the need for robust classification frameworks.
    
    These considerations motivate the development of approaches that retain coherence, broaden template coverage, and provide probabilistic assessments of candidate significance. 
    To address this, we developed a detection pipeline based on a fully coherent analysis of all detectors, incorporating both spectral and directional information. Additionally, we also derived and implemented the generalization of the matched-filter statistic (commonly used in gravitational-wave searches) to account for the Poisson statistics of photon counts. 
    This approach significantly improved sensitivity, yielding gains in SNR of a factor of $\sim2.4$–$8$ depending on the spectral properties of the burst \citep[henceforth PZV25]{perera_new_2025}. In this work, we present the results of applying our coherent Poisson matched-filter pipeline to the 13-year \textit{Fermi}/GBM dataset (2013–2025). We provide a comprehensive catalog of sub-threshold short GRBs, employing a new classification framework designed to distinguish sGRBs from SGRs, TGFs, and solar flares. By using the statistics of the full trigger dataset and timeslides analyses, we assign a class-dependent astrophysical probability $p_{\text{astro}}$ to each candidate. To further validate our low-significance candidates, we introduce a novel followup search technique using \textit{Swift}/BAT rate data to confirm GBM triggers, exploiting the instrument's sensitivity to high-energy photons even outside its coded field of view. 
    We will present a detailed analysis of the SGR population detected by our pipeline in a forthcoming paper.
    
    This paper is organized as follows: Section \ref{sec: data} describes the instruments and data products. Section \ref{sec: detection} summarizes the detection pipeline, the parameter estimation methodology, and the BAT follow-up strategy. Section \ref{sec: classification framework} details our source classification framework. In Section \ref{sec: results}, we present the results applying the pipeline to the full dataset, the statistical properties of the triggers, and comparisons with existing catalogs. We summarize and give an outlook in Section \ref{sec: summary}.
        
\section{Data and instruments} \label{sec: data}

    \subsection{Fermi Gamma Ray Burst Monitor}
        The Gamma-Ray Burst Monitor \citep{meegan_fermi_2009} on board the Fermi satellite is a GRB detector with a large field of view and wide energy coverage. It is comprised of 12 Sodium Iodide (NaI) detectors sensitive in the range of 8-1000 keV and two Bismuth Germanate detectors (BGO) covering a higher energy range of 200 keV - 40 MeV. The detectors are distributed across the spacecraft to maximize the field of view (FOV) and enable localization of bursts based on the relative flux incident on the detectors. GBM observes the whole sky unocculted by the Earth, which, due to its low Earth orbit, achieves an FOV of $8.5$ steradians. It is currently the most prolific GRB detector with $\sim 240$ detections per year. One of GBM's data products is a continuous time-tagged event (CTTE) data, which is a list of photon arrival times at each energy channel of the detector. The timing resolution is $\sim 2\;\mu$s, and the energy resolution spans 128 energy channels at $\lesssim 10\%$ resolution. Since 2012 November 27, GBM has been transmitting all the TTE data continuously. In this catalog, all the TTE data from 2013 to 2025 have been used. In our pipeline, we apply data selection criteria to exclude periods with bad time intervals and those when some detectors are inactive. All the selection criteria are described in PZV25 (Section 3.2).

    \subsection{Swift Burst Alert Telescope} \label{sec: BAT}
        The Burst Alert Telescope (BAT) on board Swift is a coded-aperture mask instrument designed to detect and localize GRBs (\cite{barthelmy_burst_2005}). It is composed of 32,768 CdZnTe detectors, sensitive in the 15-350 keV energy range, positioned 1 m below a mask, composed of 52,000 lead tiles distributed in a half-filled random pattern. When at least some part of the mask is between a point source and the detector plane, an image of the source can be obtained by deconvolving the mask from the shadow cast on the detector plane. This setup results in maximal sensitivity when a source is directly above the mask, and decreases as the source's angle increases. As a result, the FOV is angle-dependent, achieving 1.94 sr at 10\% coding and 2.85 sr at 0\% coding (\cite{krimm_swiftbat_2013}). The localization capability also depends on the coding fraction and achieves up to 4 arcminutes at full coding and no localization for zero coding. Since the BAT detection algorithms are applied only to imaged sources (\cite{fenimore_trigger_2003}), more sources are missed because they fall outside the FOV. 
        Due to its large collecting area (5200 $\text{cm}^2$), BAT's event data, comprised of arrival time, energy, and position on the detector plane, is too large to be continuously transmitted to the ground. Instead, BAT continuously transmits rate data, consisting of light curves summed over the entire detector plane in four energy channels (approx. 15–25, 25–50, 50–100, and 100–350 keV) with time bins of 64 ms, 1 s, and 1.6 s. This data product is omnidirectional; high-energy photons from sources outside the coded field of view (FOV) can penetrate the instrument shielding and register as significant rate increases, enabling temporal cross-validation of GBM triggers even without localization. In addition to rates, BAT accumulates scaled maps, which are histograms of the detector plane counts integrated over longer timescales. If a sub-threshold burst occurs within the FOV but fails to trigger the onboard logic, it may still be captured in these maps. Ground analysis can then deconvolve the mask shadow from these integrated images to recover the source position with 4 arcminutes precision.    
    
\section{Detection and parameter estimation} \label{sec: detection}
    We employ the short GRB detection pipeline developed in PZV25, which performs a coherent Poisson-based matched-filter over all Fermi/GBM detectors and energy channels to identify transient signals in the millisecond–second timescale regime. Here, we briefly summarize its main components and highlight updates introduced for this 13-year dataset \footnote{The detection pipeline and the parameter estimation codes are publicly available on GitHub: https://github.com/PeAriel/grpype}.

    \subsection{Detection pipeline overview}
    The pipeline operates on the GBM Time-Tagged Event (TTE) data in 1 ms-10 ms bins, analyzing all 14 detectors coherently. Each data segment is scanned using a stochastically placed bank of $\sim 500$ templates that span sky position and spectral shape, which provide high coverage of the parameter space available to GBM. We do this for 19 different optimally placed box templates to account for different burst durations. For every time bin and template, we compute the matched-filter statistic
    \begin{equation}\label{eq: statistic}
        \mathcal{S}_t(\theta) = \frac{\sum_{n} (d_{t,n} - b_{t,n}) \star_t \log\left(1 + \frac{AT_{t,n}(\theta)}{b_n}\right)}
        {\sqrt{\sum_{n, t} b_{t,n} \star_t \log^2\left(1 + \frac{AT_{t,n}(\theta)}{b_n}\right)}},
    \end{equation}  
    where $d_{t, n}$ is the number of photon counts in time bin $t$ in detector-energy channel index $n$, $b$ is the background, and $T(\theta)$ are the templates which are a function of the RA, DEC and Band function (\cite{band_batse_1993}) spectral parameters, which we denote as $\theta$. $A$ is the signal amplitude, which we set as the detection limit for each template based on simulations. The $\star_t$ denotes cross-correlation in the time axis. We then apply a local background correction (termed background spectral drift correction) to mitigate slow drifts in the background count rate. Candidate triggers are then subjected to a series of statistical vetoes to remove single-detector events, timing glitches, particle precipitation events, and Earth occultation steps. For the surviving triggers, we calculate the posterior distribution of sky position and spectral parameters. We apply SNR calibration to compare the significance of triggers with different durations (and hence different search volumes) as described in PZV25 (Section 4; see Eq. 16) and keep the one with the highest calibrated SNR, which is eventually reported in units of the normal distribution standard deviation. To quantify the probability of a trigger being real or a statistical fluctuation, the exact same pipeline is applied on a timeslided (\cite{was_background_2010}) dataset to empirically estimate the background trigger distribution and enable the calculation of $p_{\text{astro}}$, the probability of a trigger being real or a statistical fluctuation (\cite{venumadhav_new_2019, kapadia_self-consistent_2020}). To prevent real triggers from contaminating the timeslides distribution, we compare the SNR of triggers that occur simultaneously in both distributions. If the on-time trigger has a higher SNR, we retain it and discard the timeslide trigger. Conversely, if the timeslide trigger has a higher SNR, we discard both.
    
    \subsection{Swift/BAT followup search} \label{sec: bat followup}
        To empirically and independently validate our detections, we conducted a targeted search in the Swift/BAT rates data\footnote{Downloaded from the HEASARC Swift archive https://heasarc.gsfc.nasa.gov/FTP/swift/data} (see Sec. \ref{sec: BAT}) around our GBM trigger times. This approach is justified by the fact that BAT is sensitive to signals outside its coded mask, which are often missed due to the inability to localize them. In this work, we restrict our analysis to triggers that are sufficiently significant to exceed at least one detection threshold, either from GBM or BAT.
        
        For each of our triggers, we test whether there is a statistically significant increase in the count rate around the GBM trigger we provide. We use the likelihood ratio test to compare the signal-boxcar with the duration of the burst vs. noise-only. We assume the noise is uncorrelated Gaussian on the timescales we observe. The full derivation of the test statistic (Eq. \ref{eq: BAT statistic}) is presented in Appendix \ref{appendix: BAT statistic}. The detection algorithm works as follows: 
        \begin{itemize}
            \item We estimate the background using our rolling mean with a gap approach described in PZV25 (Section 3.3.2), where the gap duration is determined by the trigger's burst duration from our GBM detection pipeline. Using this method, we calculate the mean and standard deviation of the data, which we use as estimators for the background and it's fluctuations and denote them as $ \hat{b}_t $ and $ \hat{\sigma}_t$.
            \item We calculate the test statistic
            \begin{equation} \label{eq: BAT statistic}
                \mathcal{S}_B(t_0) = \sum_{t=t_0-\Delta/2}^{t_0+\Delta/2}\frac{(d_{t}-\hat{b}_t) \cdot T_t}{ \sigma_{t_0}\sqrt{\Delta}},
            \end{equation}
            where $\Delta$ is the duration of the burst (the length of $T$).

            \item We apply a drift correction (\cite{zackay_detecting_2021, perera_new_2025}) to account for background misestimate. This amounts to re-normalizing the test statistic
            \begin{equation}
                \mathcal{S}_{B}(t_0) \to \frac{\mathcal{S}_{B}(t_0) - \langle\mathcal{S}_{B}(t_0)\rangle}{\sqrt{\langle\mathcal{S}_{B}(t_0)\rangle^2 - \langle\mathcal{S}_{B}(t_0)^2\rangle}},
            \end{equation}
            where the expectations are estimated from the data by applying the same approach as the background estimation to the test statistic time series.
            
            \item We allow small time shifts of 1 bin between the triggers, which amounts to $\pm64$ ms for triggers with durations shorter than 1s and $\pm 1$s for triggers with durations longer than 1s. The statistic is thus
            \begin{equation} \label{eq: bat test stat}
                \mathcal{S}_B = \max_{t_0}{\mathcal{S}_B(t_0)}.
            \end{equation}
                        
        \end{itemize}
        The statistic is approximately normally distributed with a shifted mean and a slightly heavier tail due to the maximization procedure. 

\begin{figure}
            \centering
            \includegraphics[width=1\linewidth]{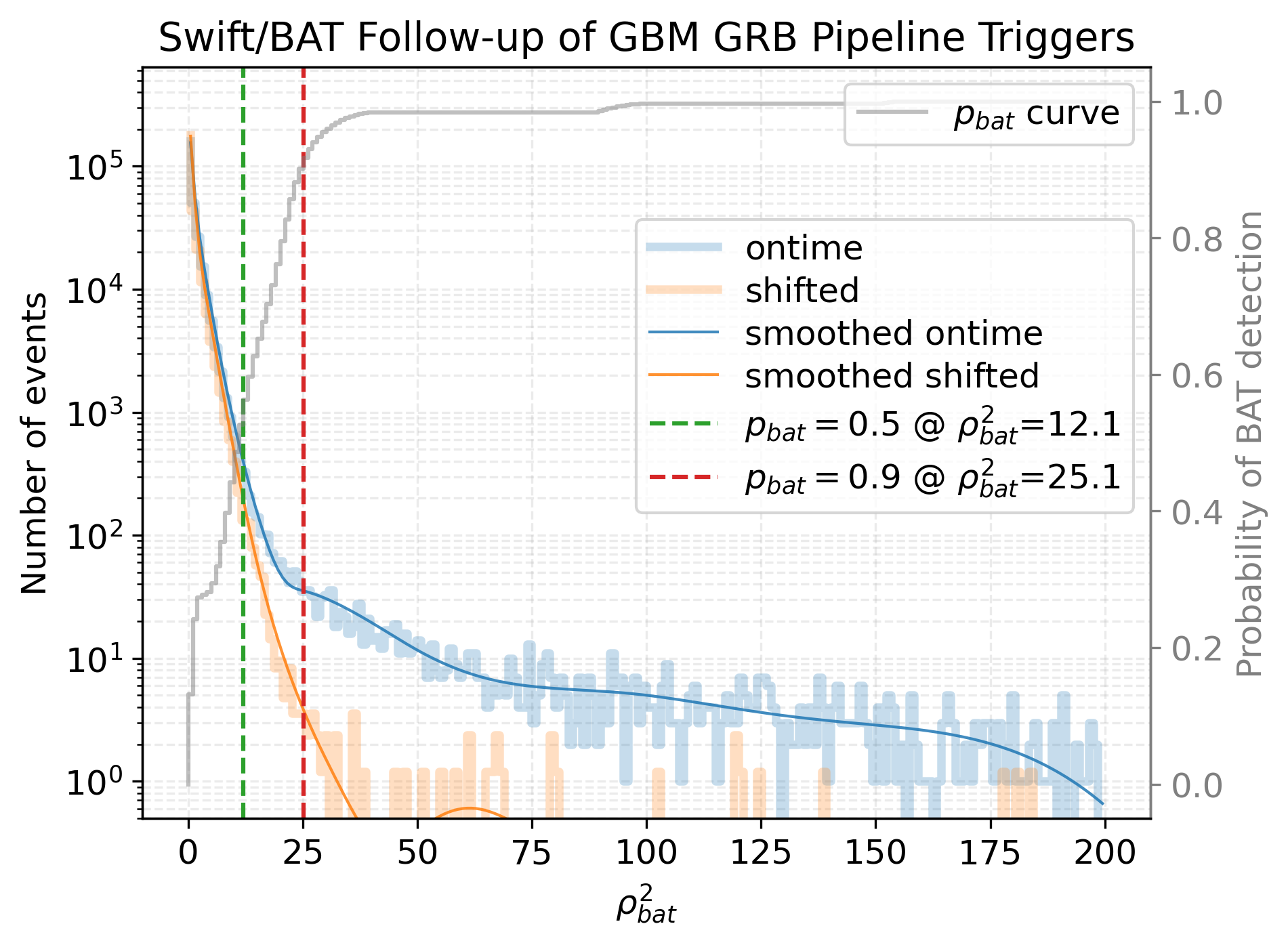}
            \caption{\textbf{Swift/BAT follow-up search trigger distribution.} The figure shows the results of the follow-up search trigger distribution, together with an empirical estimate of the background obtained from a corresponding search at time-shifted trigger times. The statistic $\rho_{\mathrm{bat}}^2$ denotes the squared test statistic (Eq. \ref{eq: bat test stat}) after drift correction. Transparent curves show the raw histograms, while solid lines indicate the smoothed distributions used to compute $p_{\mathrm{bat}}$.
            Beyond $p_{\mathrm{bat}} \approx 0.5$, the background distribution decreases significantly faster than the on-time distribution, leading to a strong excess of on-time triggers. At high probabilities ($p_{\mathrm{bat}} \gtrsim 0.9$), the shifted distribution is affected by contamination from glitches and true signals, resulting in a flattening of the inferred background. Nevertheless, the probability that a high-$p_{\mathrm{bat}}$ on-time trigger is not associated with a GBM trigger becomes negligibly small.}
            \label{fig: swift followup}
        \end{figure}

        To get an estimate of the true BAT detection probability, we also apply the same approach to the same data, shifted by 250s, which is, on the one hand, longer than the background estimation window, but small enough so that the background properties remain the same for most observations. The true BAT detection probability is calculated similarly to the $p_{\text{astro}}$ calculation described in Sec. \ref{sec: pastro}. We denote this value as $p_{\text{bat}}$. Fig. \ref{fig: swift followup} shows the resulting trigger distribution.
        This figure (and the similar Fig. \ref{fig: pastros}, discussed later) can be interpreted as follows. At low SNR${}^2$, the on-time and shifted triggers sample the same background distribution and therefore exhibit similar rates, indicating that the trigger population is dominated by noise. As the detection statistic increases, the background rate decreases rapidly, while the on-time distribution develops an excess due to astrophysical events. The point where the expected numbers of background and astrophysical triggers are equal corresponds to $p_{\rm bat}=0.5$. At progressively higher SNR${}^2$, the astrophysical population dominates, and the probability that a trigger is of astrophysical origin approaches unity. Fig. \ref{fig: swift followup} also indicates the SNR${}^2$ thresholds corresponding to $p_{\rm bat}=0.5$ and $p_{\rm bat}=0.9$, together with the complete $p_{\rm bat}$ curve.

    \subsection{Parameter estimation} \label{sec: parameter estimation}
        To enable efficient parameter estimation (PE) across a large number of triggers, we developed a new PE scheme that balances accuracy and computational cost. Our parameter estimation is based on the same optimal detection statistic used for triggering, which provides a natural likelihood function in the low-count Poisson regime. Posterior distributions can be obtained by integrating the full detection statistic (in Bayesian terminology, called the evidence) given by
        \begin{align}\label{eq: evidence}
            \mathcal{Z} = \frac{p(d|\mathcal{H}_1)}{p(d|\mathcal{H}_0)} = \int \dd \theta \; p(\theta) \frac{p(d|\theta;\mathcal{H}_1)}{{p(d|\mathcal{H}_0)}}.
        \end{align}
        Here, $\theta$ are the model parameters, which in our analysis will be a combination of the sky position $\Omega$ and spectral parameters $\xi$. For a uniform prior, the sky posterior is given by
        \begin{align} \label{eq: posterior}
            &P(\Omega | d; \mathcal{H}_1) \propto \int \dd \xi \; \frac{p(d|\Omega, \xi;\mathcal{H}_1)}{{p(d|\mathcal{H}_0)}} = \nonumber \\&\int \text{d}\xi \; \exp{\sum_n \left[d_n\log\left(1 + \frac{AT_n(\Omega, \xi)}{b_n}\right) - AT_n(\Omega, \xi)\right]},
        \end{align}
        where $\Omega$ denotes the sky position, $\xi$ the spectral parameters of the Band function, and $A$ the amplitude. The full derivation of this expression from the Poisson likelihood is presented in Appendix E of PZV25.
        In practice, we sample the evidence using Markov Chain Monte Carlo (MCMC), with the template decomposed to
        \begin{equation}
            T_n(\Omega, \xi) = \int R_n(\Omega, E)N(E, \xi) \; d E,
        \end{equation}
        where $R_n(\Omega, E)$ is the energy and sky-position dependent detector response matrix (DRM) at detector-energy channel index $n$, and $N(E, \xi)$ is the incident Photon flux having a Band function spectrum. To do the sampling we use the maximum-likelihood estimates obtained from the detection pipeline banks as initial values for the sampler. The uniform priors adopted for the Band function parameters are intentionally broader than the range typically expected for physical GRB spectra. This choice allows the phenomenological Band model to accommodate transients whose spectra are not well described by a Band function, as well as low-SNR events for which the spectral parameters are only weakly constrained. In addition, we use a set of response functions generated at a $2^{\circ}$ resolution at a fixed time and load the nearest neighbor of the sampler steps. The nearest neighbor approach was preferred over interpolating the response functions, as it is much faster and does not introduce significant errors, as the response grid generated is on a smaller scale than the systematic error in the response \citep{meegan_fermi_2009}. The choice of a $2^{\circ}$ grid therefore represents a compromise between computational efficiency and accuracy; a finer grid would provide little improvement, while a coarser grid could begin to degrade the parameter estimation. As demonstrated in PZV25 (Section 3.1; Fig. 2), using a detector response generated at an incorrect time introduces only a minor SNR loss, while generating event-specific responses for every trigger is computationally prohibitive for a large dataset of triggers. Therefore, we adopt a fixed response for our parameter estimation. Specifically, we used responses generated at 2023-10-27 14:59:49, with the reference time selected arbitrarily and verified not to coincide with any known transient activity in the GBM catalogs or with any trigger identified by our detection pipeline. After sampling, we used the best-fit parameters to refine the trigger duration estimate by scanning a dense grid of burst widths. We restrict these durations to be at most 1.5 times the original search boxcar template to avoid introducing noise (such as cosmic rays and fast-varying background) into the duration measurements. Because this parameter estimation is executed automatically over an extensive catalog of events, the sampler may occasionally fail to yield a refined posterior, most commonly due to processing issues. For any such trigger where the MCMC refinement is unsuccessful, we adopt the initial search parameters (found in the detection part) as the best available estimate for the transient.
        
        As GBM is known to have a systematic error in its sky localizations (\cite{burgess_awakening_2018, goldstein_evaluation_2020, lopez_evaluation_2024}) we performed a systematic evaluation of our localization by comparing them with \textit{Swift}/BAT jointly detected GRBs. This analysis is presented in Sec. \ref{sec: BAT comp}.
        
        \subsubsection{Results for GRB170817A}
\begin{figure*}
            \centering
            \includegraphics[width=1\linewidth]{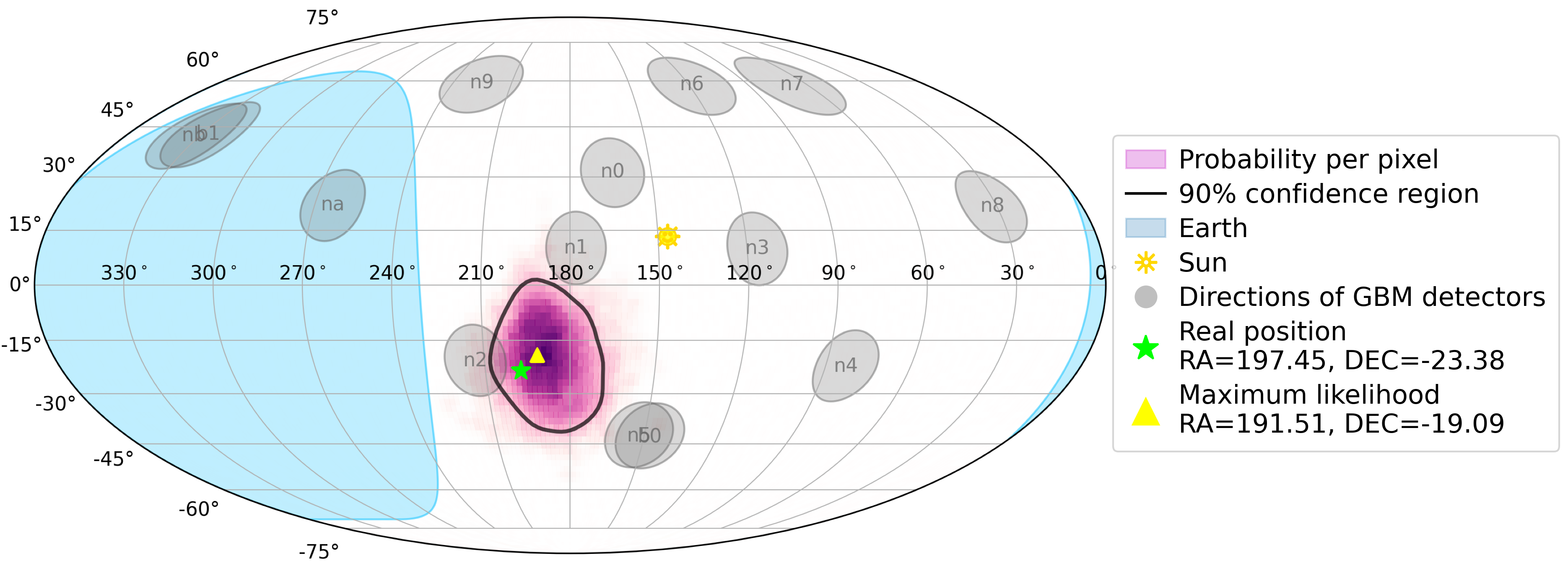}
            \caption{\textbf{The skymap of GRB170817A obtained from our parameter estimation method}. The 90\% confidence region covers approximately 1100 deg$^2$. The star marks the true GRB position, while the triangle denotes the maximum-likelihood estimate. The pointing directions of the GBM detectors are shown for illustration only and do not represent the full detector fields of view.}
            \label{fig: 170817 example skymap}
        \end{figure*}

            To illustrate the approach, as well as the pipeline's search performance, we present the results of applying the parameter estimation procedure described above to a well-known, extensively studied GRB. GRB170817A was detected by our pipeline, yielding the highest SNR for the 0.327 s box template, with a calibrated, drift-corrected SNR of $10.7\sigma$. GRB170817A was observed by GBM a few minutes before entering the South Atlantic Anomaly, which caused an increased short-term background variation. For this reason, the GBM untargeted search did not detect this event (\cite{goldstein_ordinary_2017}), and it is this increased background variation that caused the reduced SNR in our search compared to the GBM targeted search. For this duration and the initial parameters estimated from the detection pipeline, we obtained an $\sim 1100$ deg${}^2$ localization region with a small offset of the maximum likelihood estimated (MLE) position and the actual GRB position, as Fig. \ref{fig: 170817 example skymap} shows. The localization region size is much smaller than the one obtained by the GBM promptly after the event detection, which was 1800 deg${}^2$, and is similar to the refined localization they obtained by their targeted search, but did not require generating a new set of DRMs. Given the detection pipeline trigger results, the localization and parameter estimation for this burst took $\sim 10$ s. After refining the burst duration, we got a peak flux duration of 0.4 seconds. Resampling the likelihood using 0.4 s instead of the original boxcar 0.327 s gives a smaller localization region of $\sim 900$ deg${}^2$. 

    \subsection{Calculation of the astrophysical probability} \label{sec: pastro}
        The search pipeline unavoidably encounters anomalies that lead to a tail in the trigger distribution, which is not fully captured by the test statistic. To account for this, we define $p_{\text{astro}}$ as the probability that a trigger is of astrophysical origin rather than a statistical fluctuation. This probability is derived from the empirical background distribution estimated using timeslides. We compute
        \begin{equation}
            p_{\text{astro}}(\rho^2) = \frac{p(\rho^2|\mathcal{H}_1)}{p(\rho^2|\mathcal{H}_0) + p(\rho^2|\mathcal{H}_1)},
        \end{equation}
        where $\rho$ is the calibrated SNR from our detection statistic. Here, $\mathcal{H}_0$ represents the noise hypothesis and $\mathcal{H}_1$ the signal hypothesis. 
        The signal contribution $p(\rho^2 | \mathcal{H}_1)$ is estimated by subtracting the timeslides trigger-rate distribution from the on-time trigger-rate distribution. The denominator corresponds to the total on-time distribution, which includes both signal and noise contributions.
        Because the high $\rho^2$ tails of the trigger-rate distributions are subject to statistical fluctuations and exhibit non-Gaussianity, we estimate them by spline-smoothing the cumulative distribution functions and then numerically differentiating the result. This procedure yields a smooth rate density that enables a stable evaluation of $p_{\text{astro}}$.
        Finally, the timeslides triggers are accumulated over a shorter effective time-span due to additional selection cuts and data loss inherent to the sliding procedure. To ensure consistent normalization, we rescale the timeslides distribution to match the on-time distribution in the noise-dominated regime. This is achieved by multiplying the timeslides distribution by the ratio of counts in the lowest $\rho^2$ bins, where the contribution from astrophysical signals is negligible.

\section{Classification framework} \label{sec: classification framework}
        \begin{figure*}
            \centering
            \includegraphics[scale=0.7]{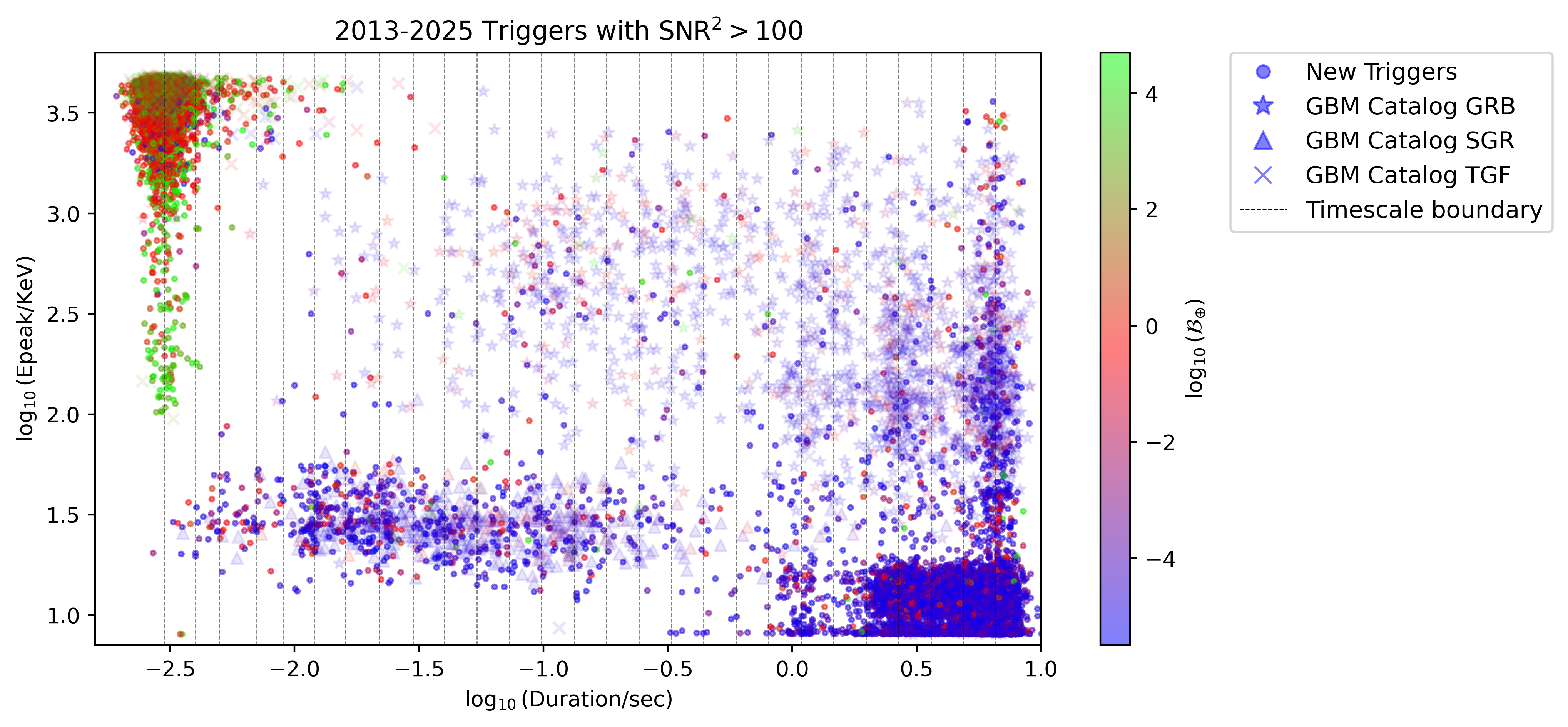}
            \caption{\textbf{Peak energy v.s. Duration trigger distribution}. The figure shows all our triggers above SNR${}^2 = 100$ - a representative value for credible detections across all classes. Triggers that were also classified by the GBM catalog are marked by different shapes as indicated in the legend. A small jitter has been added to the duration axis to break the grid structure, which is indicated as dashed vertical lines. The peak energy is the median of the marginalized posterior. There are different, distinct regions in this figure that can be attributed to different source classes.}
            \label{fig: E-t plane}
        \end{figure*}
    
\begin{figure*}
            \centering
            \includegraphics[scale=0.6]{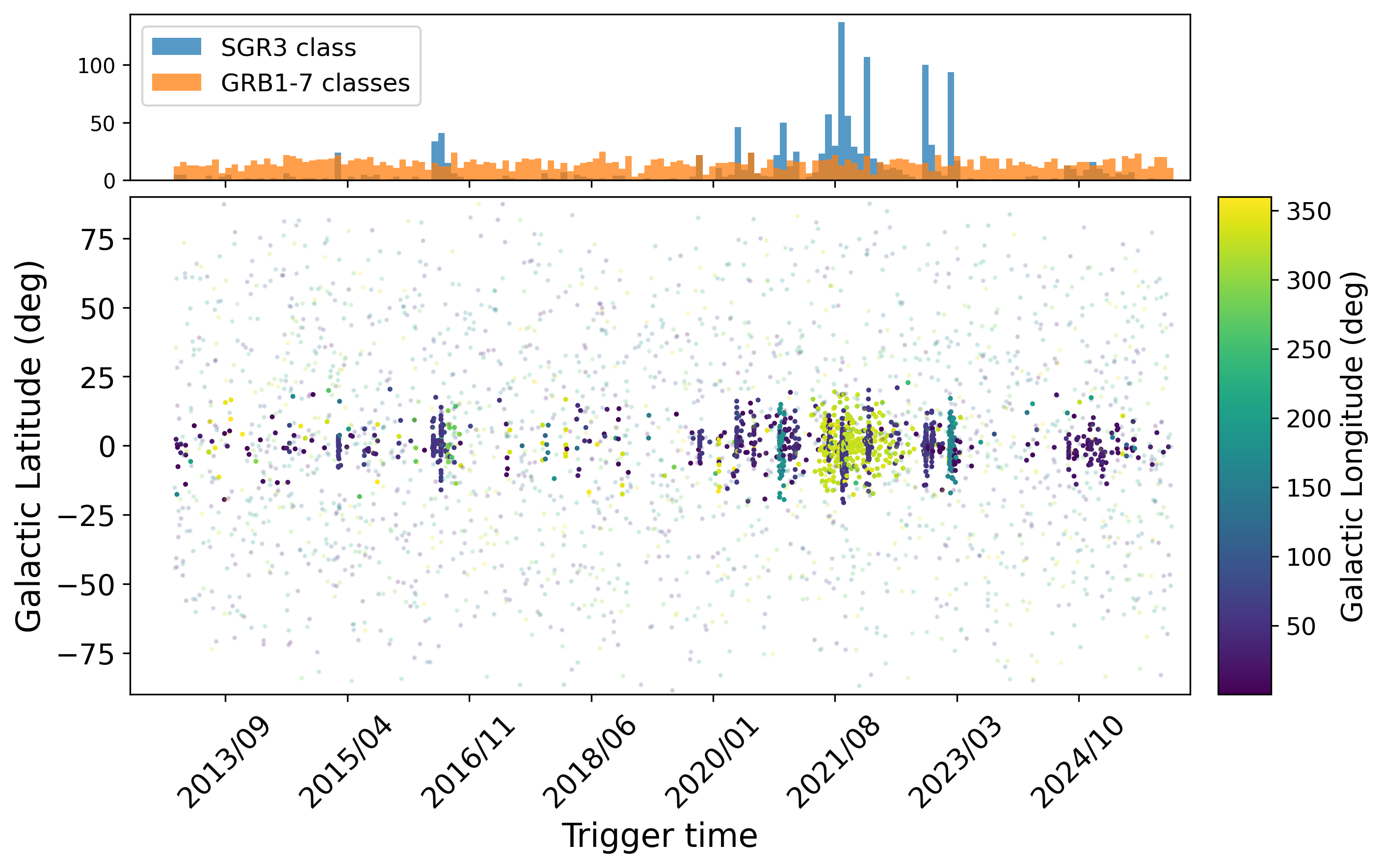}
            \caption{\textbf{Burst occurrence times of Magnetars and GRBs}. The top panel shows a histogram of the burst occurrence times for triggers in the SGR3 class (see Fig. \ref{fig: tree} for the class definitions), whose events likely correspond to magnetar bursts, and for triggers in the GRB1-7 classes. The bottom panel shows individual triggers in the Galactic latitude-trigger time plane, color-coded by the galactic longitude, where the opaque dots are from the SGR3 class and the transparent dots are for the GRB classes. Spatio-temporal clustering of the Magnetar-like bursts is clearly observed across multiple timescales, whereas the GRB-like events are uniformly distributed in both time and space.}
            \label{fig: magnetar storms}
        \end{figure*}

    Our extensive sample of GBM triggers provides a means to uncover underlying patterns that aid in distinguishing between different astrophysical sources, despite the limited localization accuracy of GBM. The combined information from the peak energy, trigger duration, trigger time, and the sky position (including the Earth and Sun Bayes factors) is used to place selection cuts in order to classify the triggers. Nevertheless, contamination between the different source classes - GRBs, SGRs, TGFs and SFs is inevitable. Fig. \ref{fig: E-t plane} shows the trigger distribution in the $E_{\text{peak}}$ - duration plane, color coded by the Earth Bayes factor (a statistical measure to quantify the probability of a trigger originating from the Earth or any other direction, see PZV25 Section 3.4.5 and Appendix F) and GBM's catalog triggers are marked with different shapes.
    One can see clustering of triggers in different areas, indicating different source classes. TGFs, which have very short durations and high peak energy, are clustered at the top left corner, though tails extending to higher timescales and lower energies are present. The color scale also illustrates the loss of directional information at higher energies. Many high-energy triggers have Earth Bayes factors close to unity, reflecting the reduced directional sensitivity of the GBM detectors in this energy regime. The horizontal line at the bottom right (longer softer emission) is GBM's lower energy cutoff, where triggers in this region are from solar flares. The blob above this line is comprised mainly of X-ray binary flares, and the gap towards the shorter end is due to our discrete duration template grid. The softer, shorter emission is due mainly to Magnetar bursts, but X-ray binary flares may also be present in this regime. The presence of these distinct populations is evident in the temporal distribution of the triggers. As shown in Fig. \ref{fig: magnetar storms},
    triggers in the softer, short-duration, Magnetar regime (SGR3 class, see Sec. \ref{sec: trigger classification} and Fig. {\ref{fig: tree}}) often occur in dense temporal clusters or 'burst storms', contrasting with the more uniform temporal distribution of GRB-like events. While a formal attempt to identify and associate specific Magnetar sources via temporal and spatial clustering will be presented in an accompanying SGR burst catalog paper, the clear presence of these heterogeneous populations within the data strongly motivates the need for a robust classification scheme. The middle region in Fig. \ref{fig: E-t plane} is where we expect to find most of the GRBs, as indicated also by the agreement with the GBM catalog. 
                
    \subsection{Selection cuts for trigger classification} \label{sec: trigger classification}
\begin{figure*}
            \centering
            \includegraphics[scale=0.85]{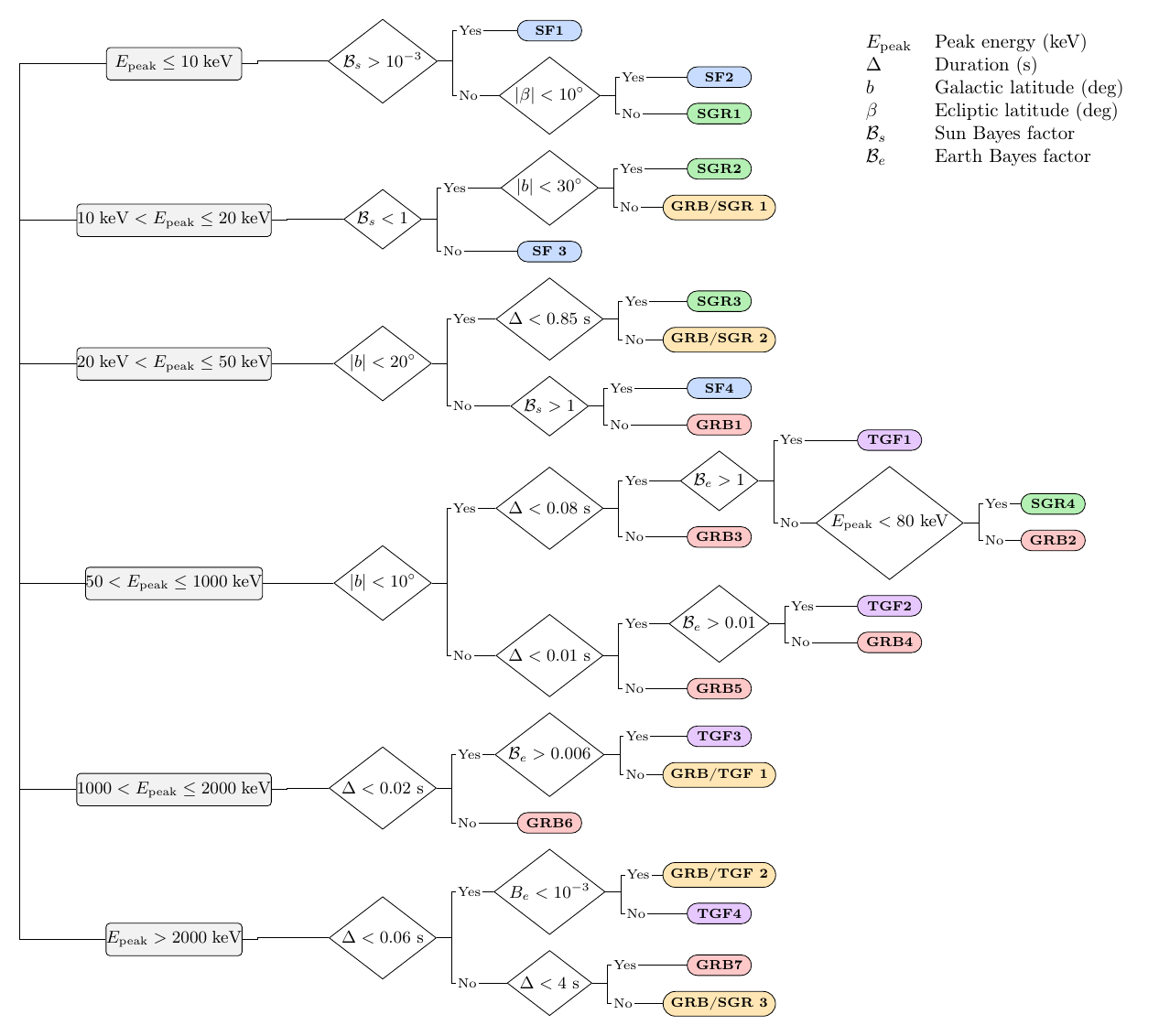}
            \caption{\textbf{Classification Scheme for our triggers}. This tree outlines the decision process we use to classify triggers in our catalog. }
            \label{fig: tree}
        \end{figure*}

         We use the information discussed above to construct a decision process for classifying triggers. The analysis is divided into six peak energy ranges: below 10 keV, 10-20 keV, 20-50 keV, 50-1000 keV, 1000-2000 keV, and above 2000 keV. We further exclude triggers for which the MLE of the Band function high-energy index $\beta$ (\cite{band_batse_1993}) exceeds -1.5, since such values may lead to nonphysical flux estimates. This threshold, together with the subsequent classification criteria, is motivated by a manual exploration of the parameter space using pair plots, similar to those shown in Fig. \ref{fig: E-t plane}, for the relevant parameters. A schematic representation of the classification procedure is presented in Fig. \ref{fig: tree}.
         
         We applied this decision tree uniformly to all triggers in our sample, using the median parameter values derived from our parameter estimation to evaluate the conditions at each node. Each leaf of this decision tree corresponds to a class of triggers identified in the data. We emphasize that the numerical labels assigned to different phenomena (e.g., GRB 1 and GRB 2) serve solely as an observational convenience and do not imply physically distinct subclasses; they are used to group triggers and assign probabilities of being astrophysical signals or noise based on their observed properties from the pipeline and parameter estimation stage. Classes that include two transients, e.g., GRB/SGR, are likely to have a high contamination level, and manual examination is required to identify individual triggers.


\section{Results and discussion: The 13-Year Catalog} \label{sec: results}
\begin{figure*}
    \centering
    \includegraphics[scale=0.39]{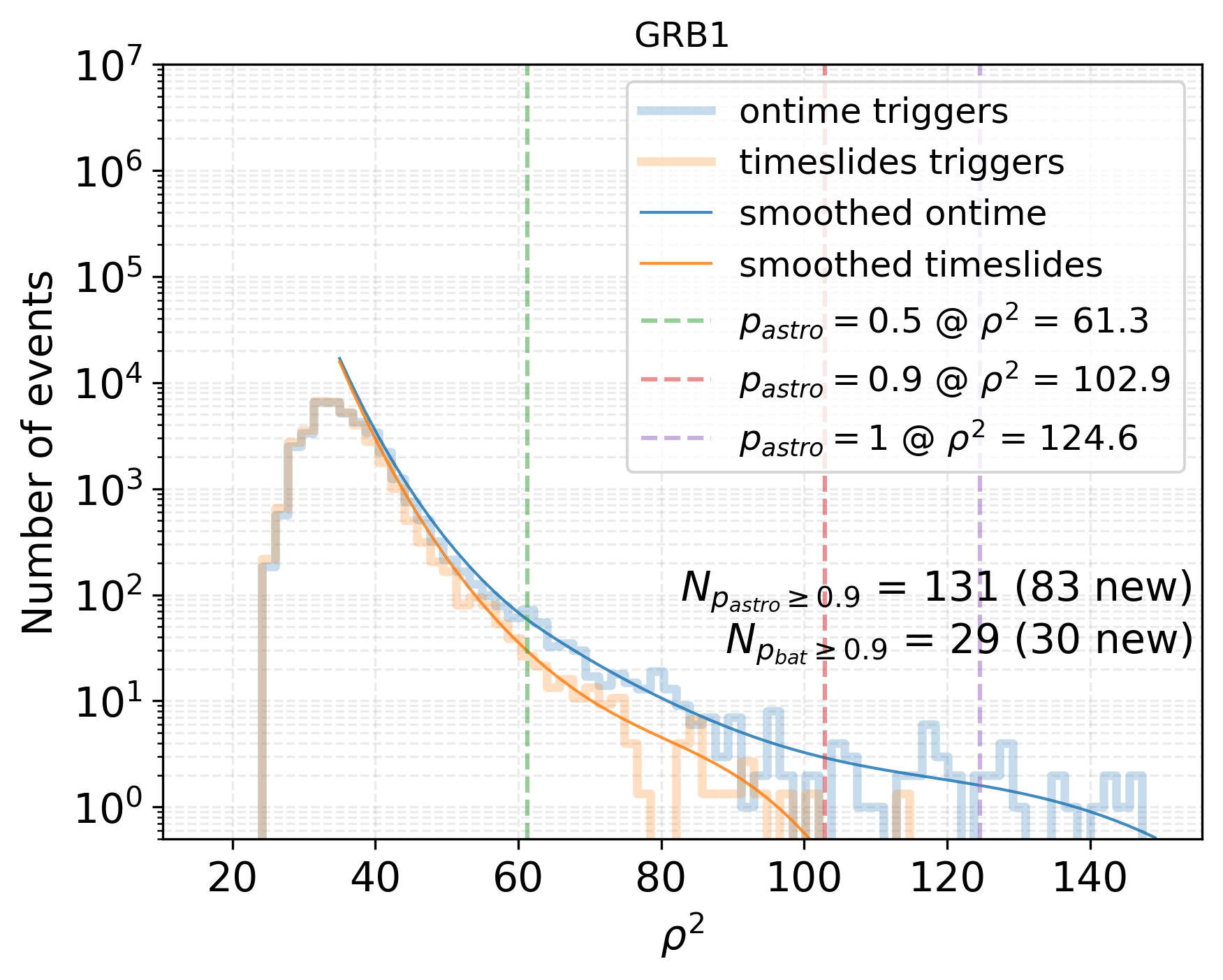}
    \includegraphics[scale=0.39]{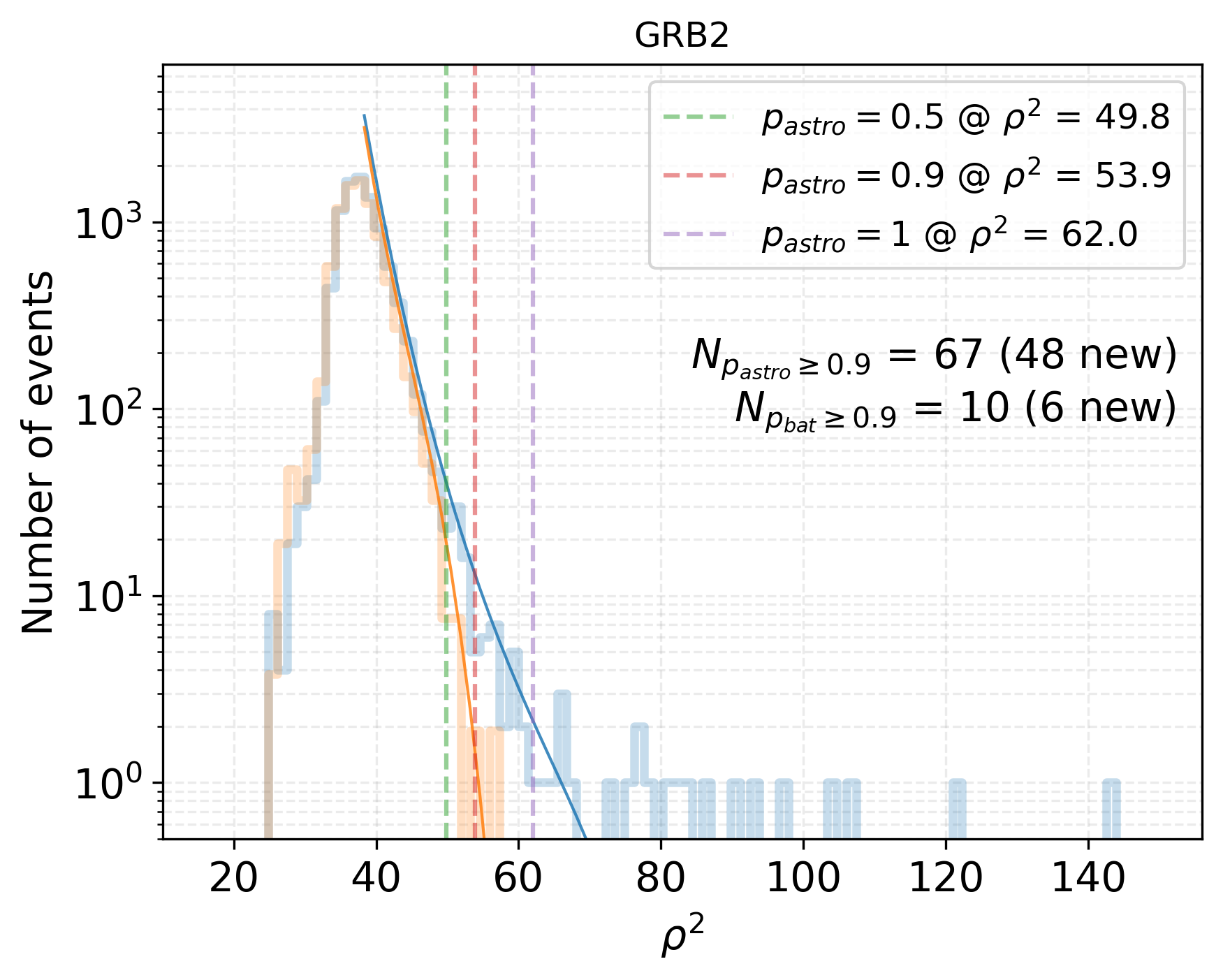}
    \includegraphics[scale=0.39]{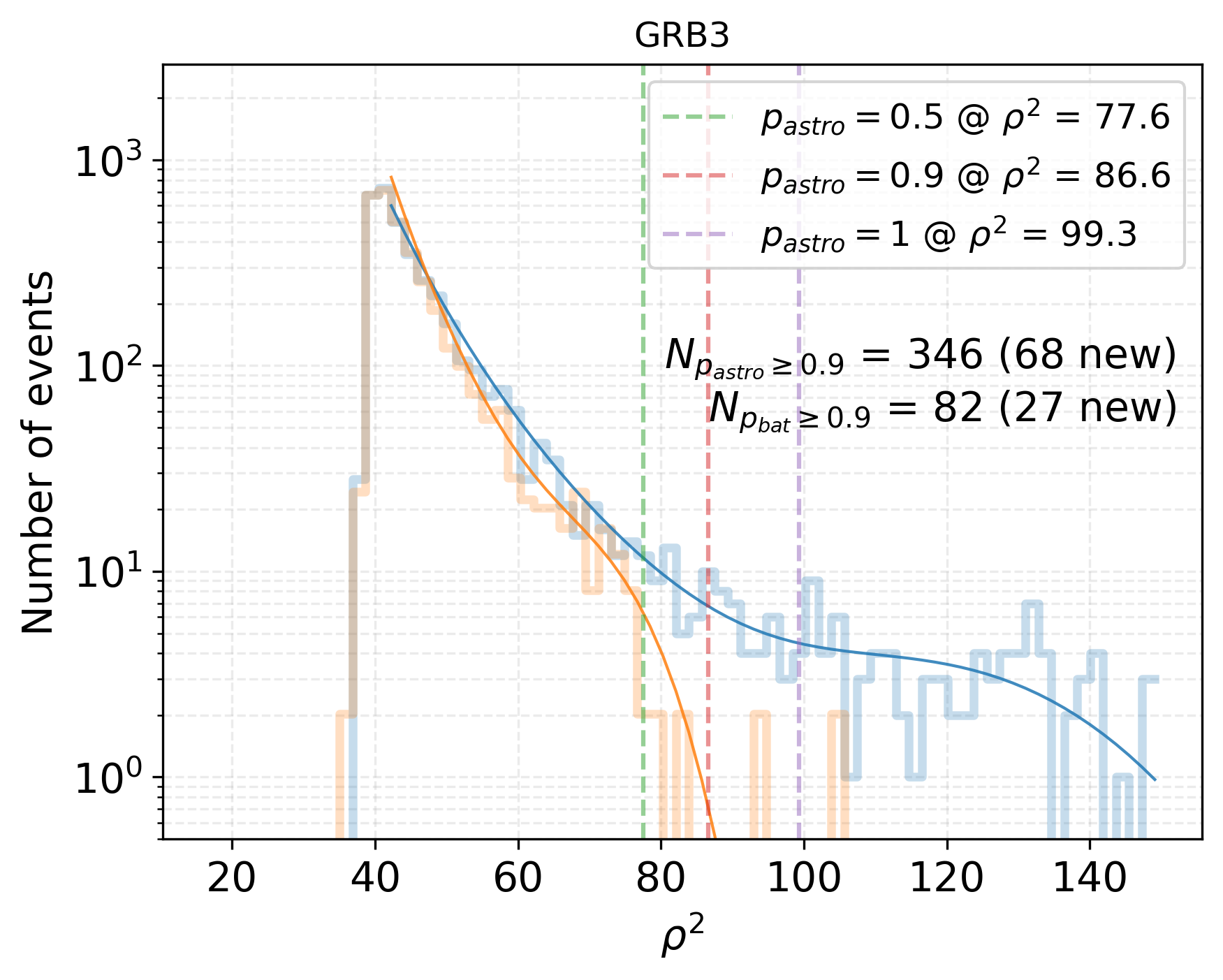}
    \includegraphics[scale=0.39]{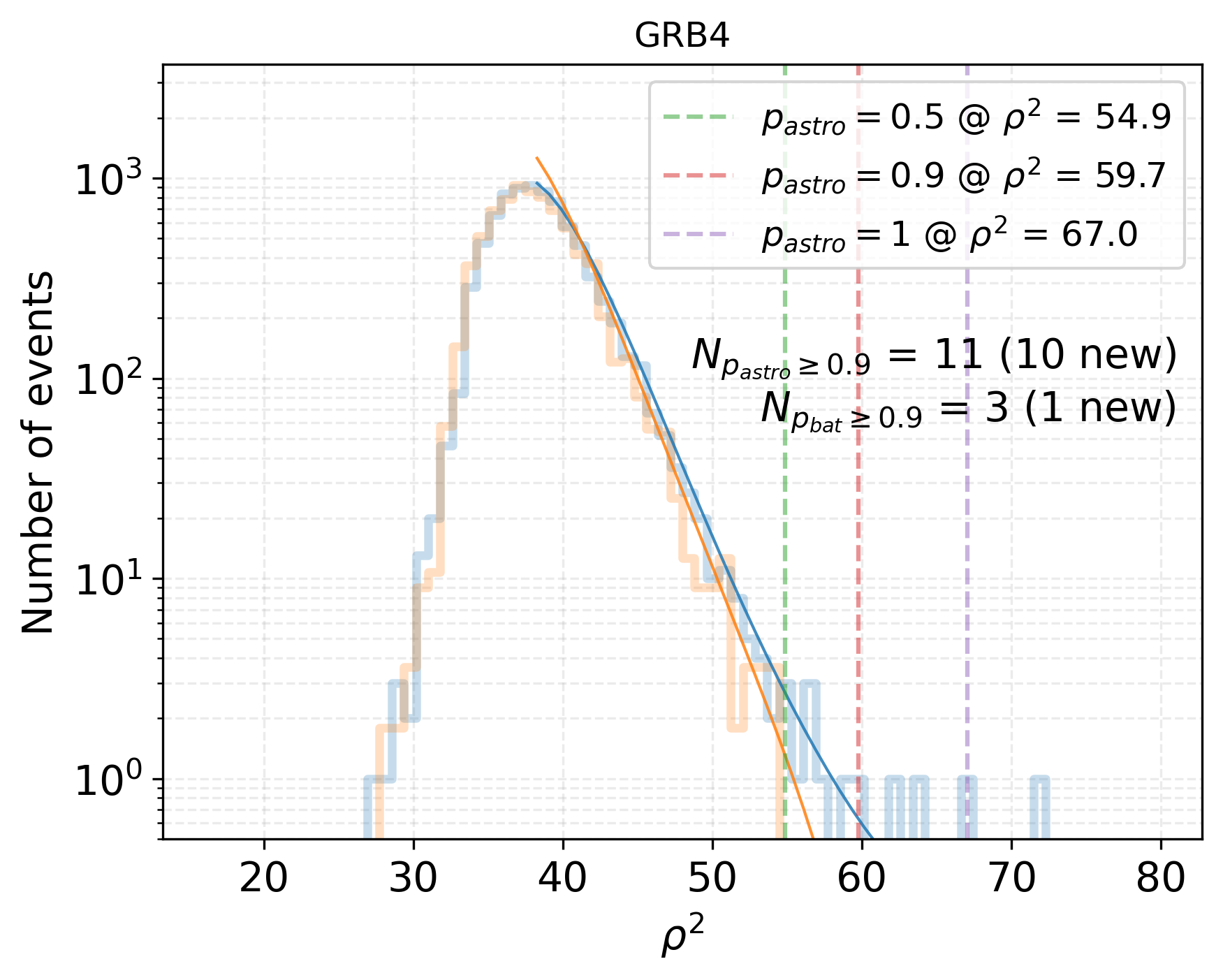}
    \includegraphics[scale=0.39]{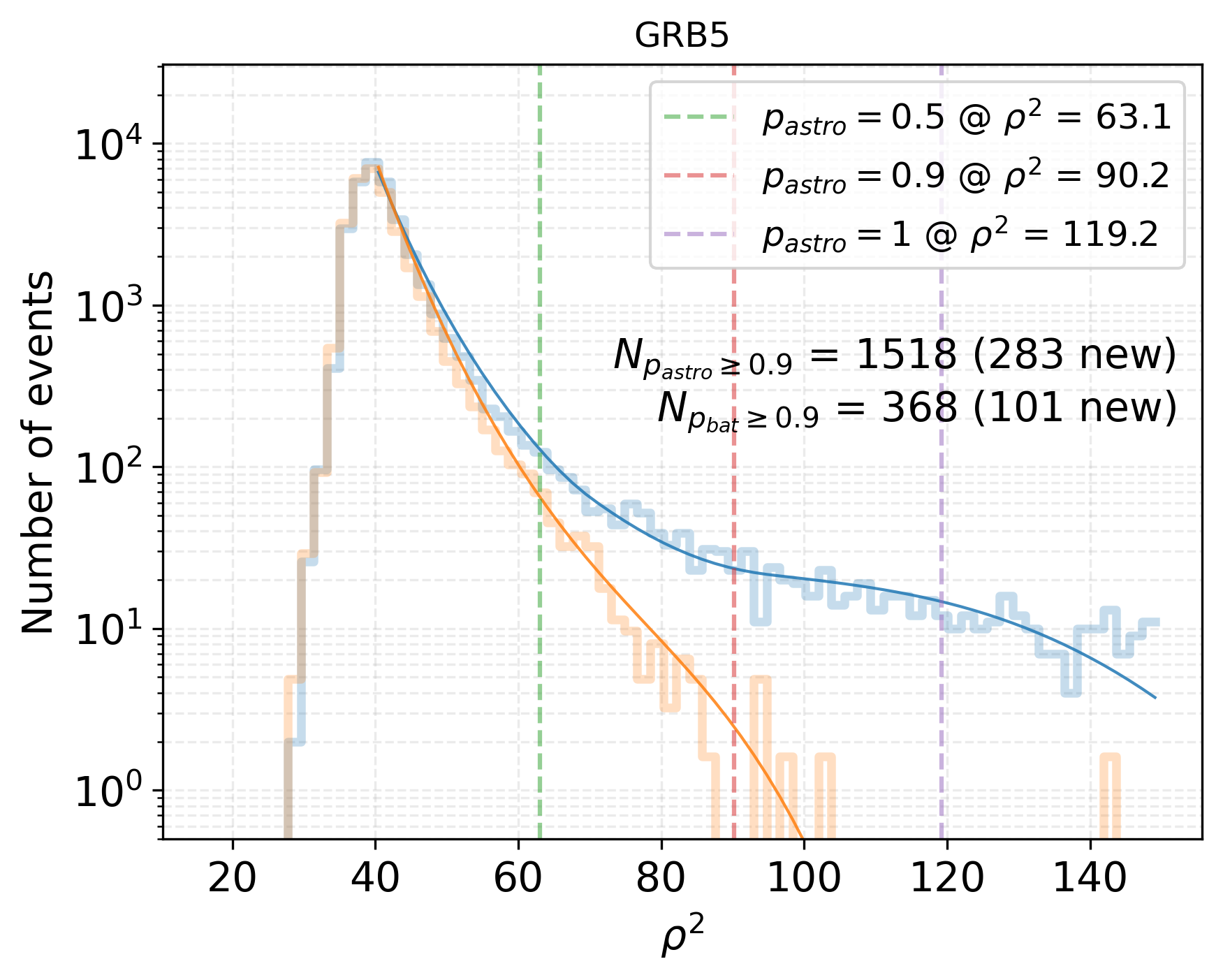}
    \includegraphics[scale=0.39]{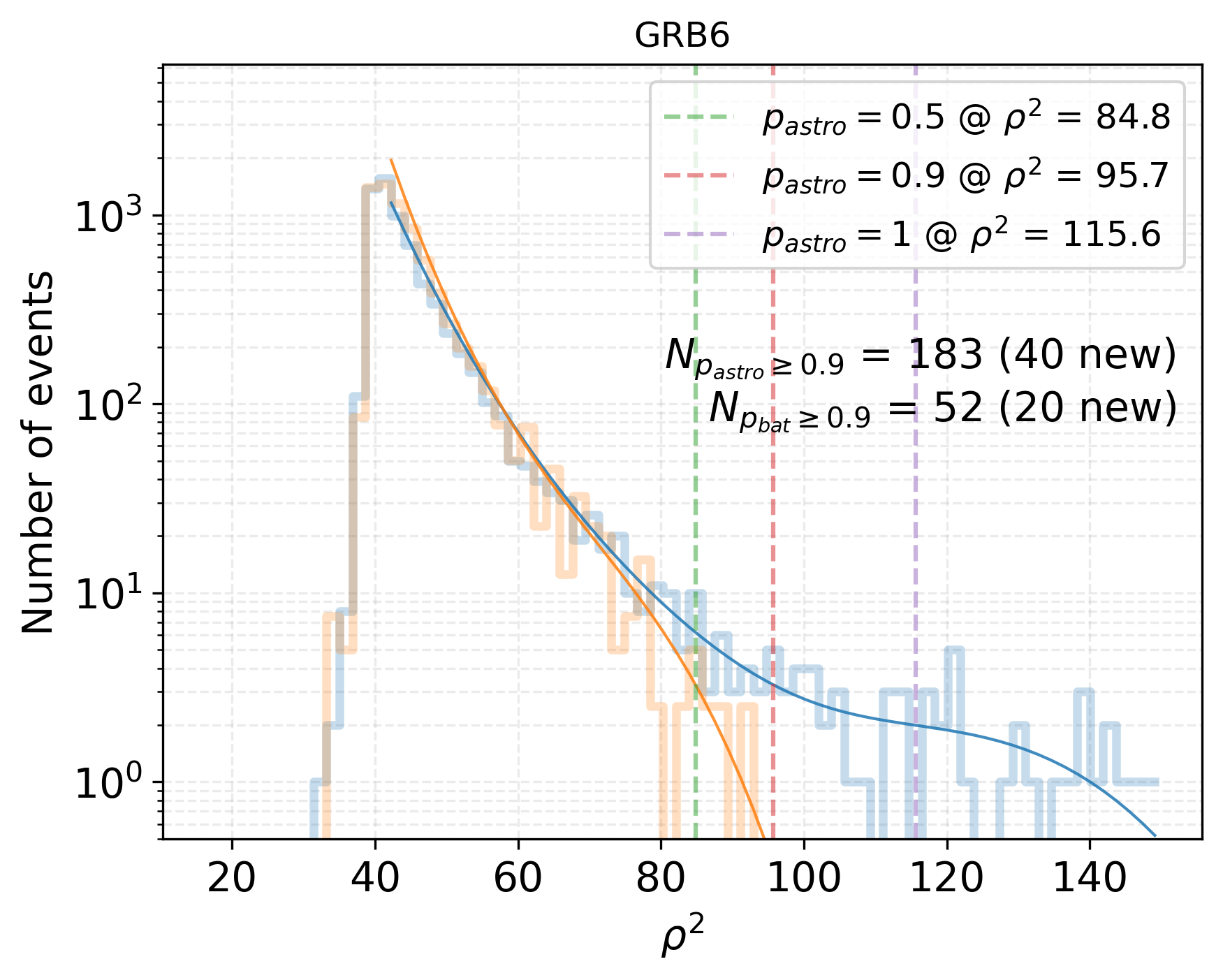}
    \includegraphics[scale=0.39]{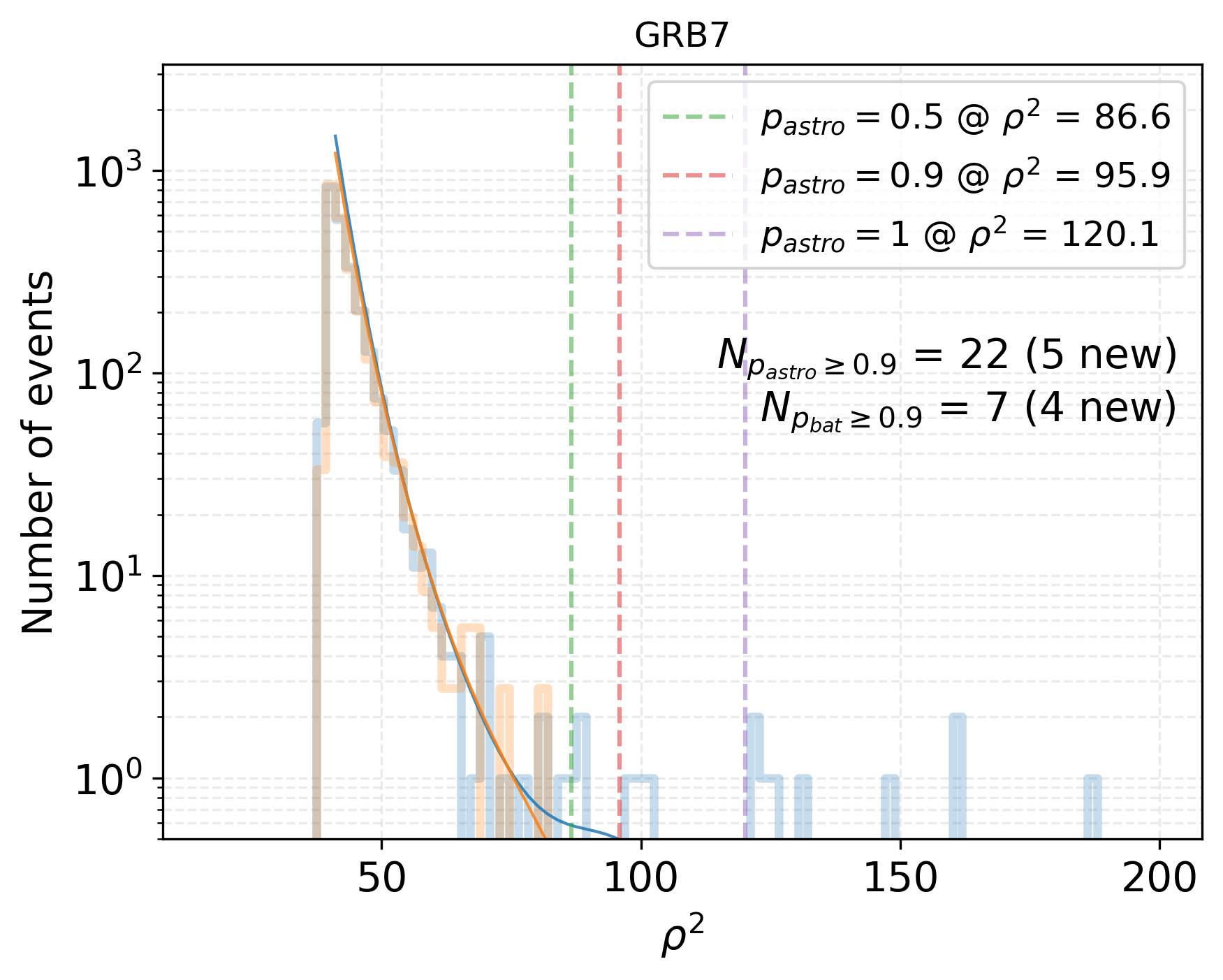}
    \caption{\textbf{Trigger distribution for the different GRB classes.} The panels show the on-time vs. timeslides distributions and the smoothed distributions used in the $p_{\text{astro}}$ calculation. We also overlay the counts of triggers with $p_{\text{astro}}\geq0.9$ and $p_{\text{bat}}>0.9$. The "new" label indicates the number of triggers not found in the current GBM catalog.}
    \label{fig: pastros}
\end{figure*}

    The complete catalog of triggers, along with the derived parameters and classification information described in the previous sections (such as $p_{\text{astro}}$, $p_{\text{bat}}$, etc.), is publicly available online in a Zenodo repository: \href{https://doi.org/10.5281/zenodo.20430377} {10.5281/zenodo.20430377} \citep{perera_catalog_2026}. Also included in this catalog are the posterior samples of triggers with SNR${}^2\geq50$. A detailed description of the catalog columns is provided in Appendix \ref{sec:appendix_catalog}.
    Because the SNR distribution varies across parameter space, each class defined by our selection cuts exhibits a different trigger distribution. This, in turn, leads to different detection thresholds in SNR for each source class. We present the trigger distributions for the 7 GRB classes in Fig. \ref{fig: pastros}.

    Together with the timeslides triggers, these are used to derive the $p_{\text{astro}}$ distribution for each class. This figure also indicates the number of events with $p_{\text{astro}}\geq0.9$ and $p_{\text{bat}}\geq0.9$. For some classes (e.g., GRB 5), there is a long tail of timeslide triggers, resulting in a high SNR$^2$ value for $p_{\text{astro}}=1$. Moreover, even for lower SNR$^2$ values, the number of candidates sometimes outnumbers the expected background from timeslides. For these reasons, we adopt $p_{\text{astro}}\geq 0.9$ as the threshold for our analysis below.

    Throughout this work, we define a trigger as \emph{previously unknown} (or new) if it cannot be associated with any GRB in the GBM catalog. To determine this, we compare each trigger in our catalog with every GBM GRB and consider the events associated if the trigger falls within the published $T_{90}$ interval of the corresponding GBM event. To account for the different duration definitions used by our pipeline and the GBM catalog, we conservatively extend this association window by an additional $\pm20$ seconds.
        
    Within the GRB classification, our pipeline yields 1950 triggers (341 previously unknown) with $p_{\text{astro}}=1$, and 2292 triggers (549 previously unknown) with $p_{\text{astro}}\geq 0.9$. Assuming all of these events are short GRBs, even our most conservative sample ($p_{\text{astro}}= 1$) reproduces the $\sim50\%$ increase in short GRB detections reported by PZV25. Conversely, if the pipeline's ratio of short to long GRBs remains unconstrained, a rough estimate that 50\% of the $p_{\text{astro}} \geq 0.5$ triggers are genuine results in a $\sim30\%$ increase in the total number of combined short and long GRB detections.
        
\begin{figure*}
        \centering
        \includegraphics[scale=0.5]{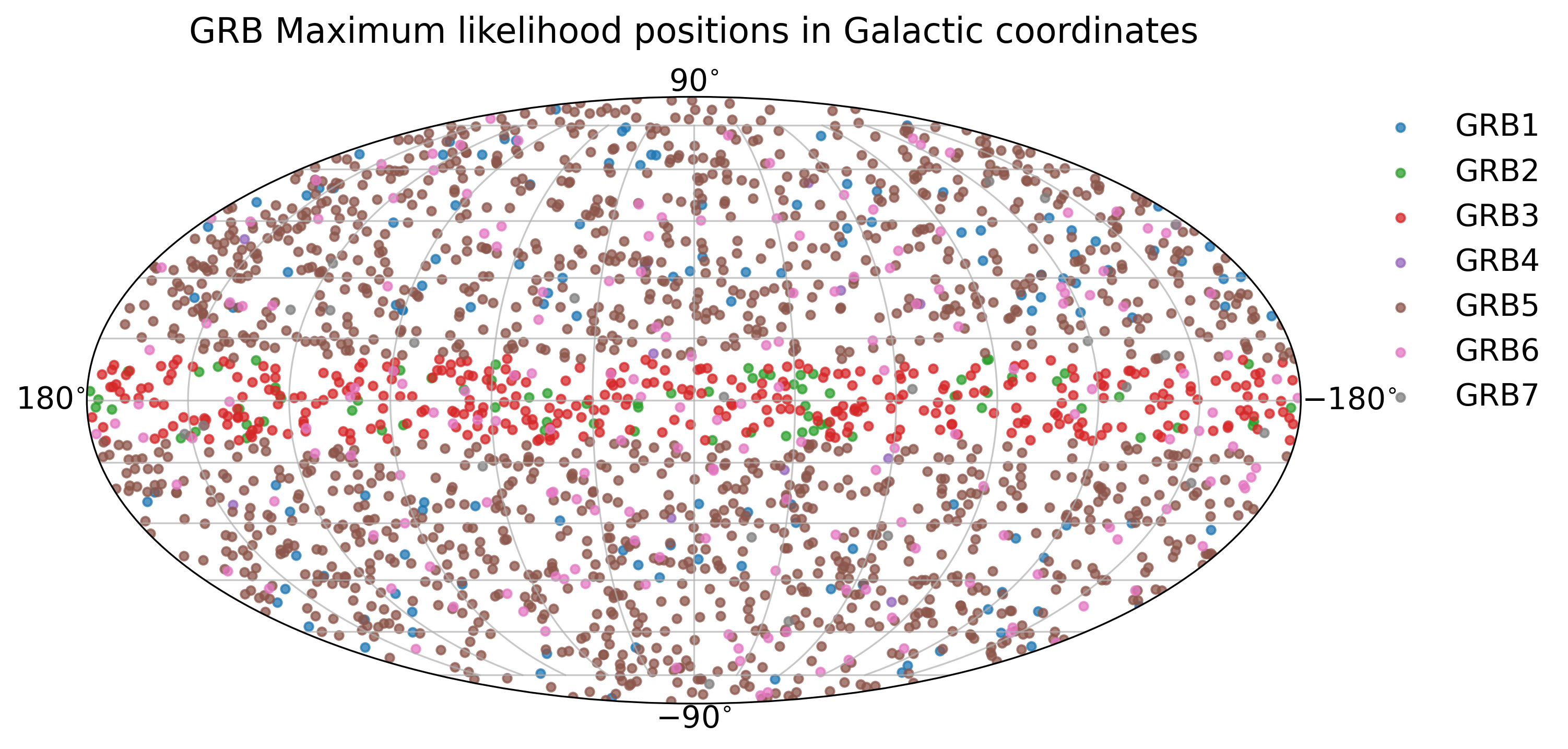}
        \caption{\textbf{Sky distribution of GRB candidates.} The figure displays the MLE sky positions in Galactic coordinates for triggers classified as GRBs with $p_{\text{astro}}\geq 0.9$. The combined distribution of the 7 GRB classes appears isotropic, consistent with an extragalactic origin. Notably, the "GRB2" and the "GRB3" classes, which consist of triggers located near the Galactic plane but with spectral and temporal properties that make them unlikely to be SGRs, show no significant spatial clustering. This lack of anisotropy further strengthens their association with GRBs rather than Galactic SGRs.}
        \label{fig:sky_distribution}
    \end{figure*}

        
    We first examine the sky distribution of our candidates to verify their extragalactic nature.
    As shown in Fig. \ref{fig:sky_distribution},
    The combination of all 7 GRB classes produces a uniform sky distribution as expected for cosmological sources. By construction, events near the Galactic plane belong to the "GRB2" and "GRB3" classes. The "GRB2" class is characterized by peak energies above 80 keV and durations shorter than 80 ms, whereas the "GRB3" class spans peak energies between 50 keV and 1 MeV and durations longer than 80 ms. The 50 and 80 keV thresholds, shown explicitly in the decision tree (Fig. \ref{fig: tree}), were chosen from the duration-peak energy distribution (Fig. \ref{fig: E-t plane}), where they provide an empirical separation between the bulk of the magnetar burst population and the GRB candidates. Although these events are located near the Galactic plane, their spectral and temporal properties make them unlikely to be SGRs. In addition, they do not form clusters, which are expected for persistent sources. Instead, they are widely spaced in longitude and effectively fill the plane region of the map, reinforcing the overall uniform sky distribution. This behavior, which arises from our selection cuts and classification scheme, increases our confidence that the GRB classes are not significantly contaminated by Galactic sources.    
    In the following, we compare our catalog to the GBM and Swift/BAT GRB catalogs.
    
    \subsection{Comparison with the GBM GRB catalog} \label{sec: gbm catalog comp}
        For each of our triggers, we searched for a temporally coincident burst in the continuously updated Fermi/GBM Burst Catalog available through the Fermi Science Support Center\footnote{\url{https://heasarc.gsfc.nasa.gov/w3browse/fermi/fermigbrst.html}} (hereafter, the GBM catalog; \cite{von_kienlin_fourth_2020, gruber_fermi_2014, von_kienlin_second_2014, bhat_third_2016}). For this trigger recovery comparison, we use a symmetric matching window of $\pm 6.573$ sec, corresponding to the duration of the longest search template employed by our pipeline. This window is sufficiently large to account for differences between the GBM $T_{90}$ duration and our peak-flux duration estimate, while remaining small enough to minimize the probability of misassociating unrelated events. We use these associations to compare our trigger classifications with those reported in the GBM catalog. There are 857 GRBs in GBM's catalog with a duration less than our maximal boxcar template of 6.573 s. Our pipeline recovers 746 GBM GRBs out of which 731 are classified as GRBs and 700 have $p_{\text{astro}}\geq 0.9$. The main reasons for trigger losses are our vetoes and the preprocessing disposal of the edges of data files. 
        
        Although our pipeline is not optimized to search for long GRBs, we repeat the above analysis, including GBM bursts with durations longer than 6.573 sec, for completeness. This comparison accounts for the fact that our duration estimate corresponds to the peak-flux duration rather than $T_{90}$ (see Section \ref{sec: temporal} and Fig. \ref{fig: duration comp}). For this matching, we use a broader matching window of $\pm20$ s to account for the larger offset that may exist between the GBM trigger time and the peak-flux duration identified by our pipeline. Using this criterion, we identify and classify 1990 triggers (1618 with $p_{\text{astro}}>0.9$) as GRBs out of the 3122 GBM GRBs. We also note that this comparison is only a rough estimate, since for long-duration GRBs, we need a longer comparison baseline than 20 sec, but if we increase this window, we risk the association of unrelated events.
        
\begin{figure}
            \centering
            \includegraphics[width=\columnwidth]{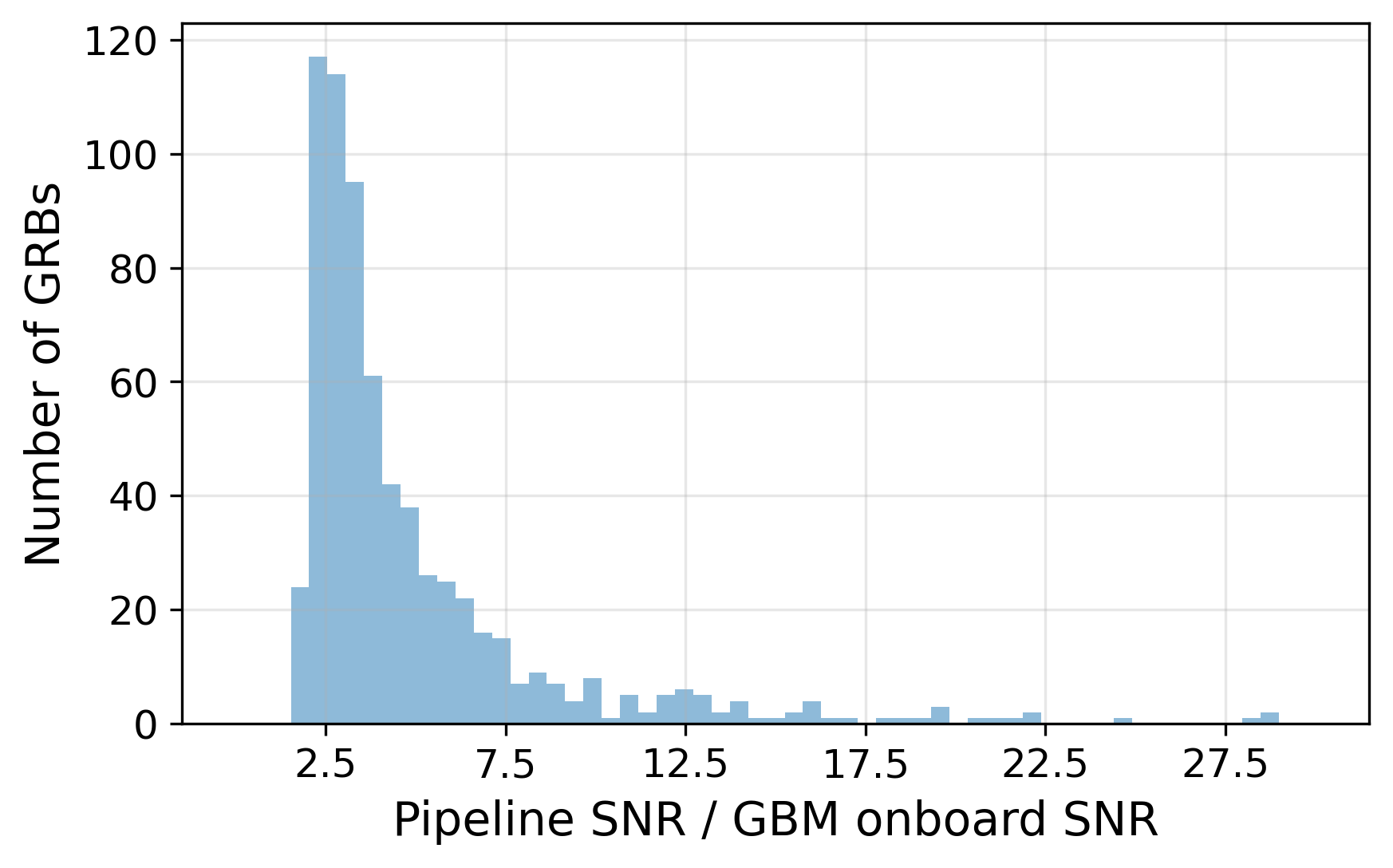}
            \caption{\textbf{Detection SNR comparison}. This figure shows the ratio of the detection SNR of our pipeline compared with the detection SNR of the GBM onboard triggering algorithm. Our pipeline achieved a significant increase in SNR based on the burst spectral properties.}
            \label{fig: snr comp}
        \end{figure}

        A comparison of our classification for $p_{\text{astro}}\geq 0.9$ triggers and GBM's classification is presented in Table \ref{tab: gbm comparison} and a comparison of our detection SNR and the GBM onboard triggering algorithm SNR is presented in Fig. \ref{fig: snr comp}. Since both our calibrated SNR and the onboard SNR determine the detection significance and are reported in units of the standard normal distribution, they provide a common measure for comparing the sensitivity of the two searches. The figure illustrates the increase in detection significance achieved by our coherent search, which explains the larger number of detected events. From Table. \ref{tab: gbm comparison} one can see that there is a broad agreement between our classifications and GBM's classifications.

        \begin{table*}
            \centering
            \caption{Comparison between our pipeline's $p_{\text{astro}}\geq 0.9$ triggers and GBM's classification.}
            \label{tab: gbm comparison}
            \begin{tabular}{lrrrrrrrrr}
            \hline
            Pipeline Class &
            $N_{p_\text{astro}\geq 0.9}$ &
            $N_{\text{GBM}}$ &
            lGRB &
            sGRB &
            TGF &
            SGR &
            SFLARE &
            UNCERT &
            Other \\
            \hline
GRB1            & 131          & 48       & 43    & 3     & 0     & 1     & 0     & 0     & 1     \\
GRB2            & 67           & 19       & 6     & 13    & 0     & 0     & 0     & 0     & 0     \\
GRB3            & 346          & 278      & 231   & 42    & 0     & 1     & 1     & 1     & 2     \\
GRB4            & 11           & 1        & 0     & 1     & 0     & 0     & 0     & 0     & 0     \\
GRB5            & 1518         & 1235     & 894   & 314   & 1     & 4     & 3     & 16    & 3     \\
GRB6            & 183          & 143      & 64    & 74    & 1     & 0     & 1     & 2     & 1     \\
GRB7            & 22           & 17       & 10    & 7     & 0     & 0     & 0     & 0     & 0     \\
(GRB/SGR)1      & 23           & 0        & 0     & 0     & 0     & 0     & 0     & 0     & 0     \\
(GRB/SGR)2      & 151          & 26       & 15    & 0     & 0     & 9     & 0     & 0     & 2     \\
(GRB/SGR)3      & 4            & 3        & 3     & 0     & 0     & 0     & 0     & 0     & 0     \\
(GRB/TGF)1      & 24           & 3        & 1     & 1     & 0     & 0     & 0     & 1     & 0     \\
(GRB/TGF)2      & 379          & 68       & 4     & 1     & 59    & 0     & 0     & 4     & 0     \\
TGF1            & 45           & 3        & 0     & 1     & 1     & 1     & 0     & 0     & 0     \\
TGF2            & 256          & 4        & 0     & 2     & 2     & 0     & 0     & 0     & 0     \\
TGF3            & 1094         & 7        & 0     & 1     & 4     & 0     & 0     & 2     & 0     \\
TGF4            & 9079         & 911      & 10    & 0     & 861   & 2     & 3     & 19    & 16    \\
SGR1            & 293          & 6        & 6     & 0     & 0     & 0     & 0     & 0     & 0     \\
SGR2            & 6831         & 62       & 13    & 0     & 5     & 39    & 0     & 1     & 4     \\
SGR3            & 1397         & 426      & 3     & 1     & 0     & 405   & 4     & 5     & 8     \\
SGR4            & 47           & 8        & 0     & 0     & 0     & 7     & 0     & 1     & 0     \\
SF1             & 2671         & 28       & 6     & 0     & 3     & 1     & 17    & 1     & 0     \\
SF2             & 92           & 2        & 1     & 1     & 0     & 0     & 0     & 0     & 0     \\
SF3             & 434          & 16       & 1     & 0     & 0     & 0     & 14    & 1     & 0     \\
SF4             & 21           & 1        & 0     & 0     & 0     & 0     & 1     & 0     & 0     \\
            \hline
            \end{tabular}
            \medskip\\
            {\footnotesize \textit{Note.}
                Associations are based on the presence of a GBM catalog event as described in Sec. \ref{sec: gbm catalog comp}.
                $N_{p_\text{astro}\geq0.9}$ denotes the number of pipeline triggers in each class with astrophysical probability $p_\text{astro} \geq 0.9$.
                $N_{\text{GBM}}$ denotes the number of those triggers that are associated with a GBM catalog classification.
            }
        \end{table*}
        
        \subsubsection{Temporal properties} \label{sec: temporal}
\begin{figure}
                \centering
                \includegraphics[width=\columnwidth]{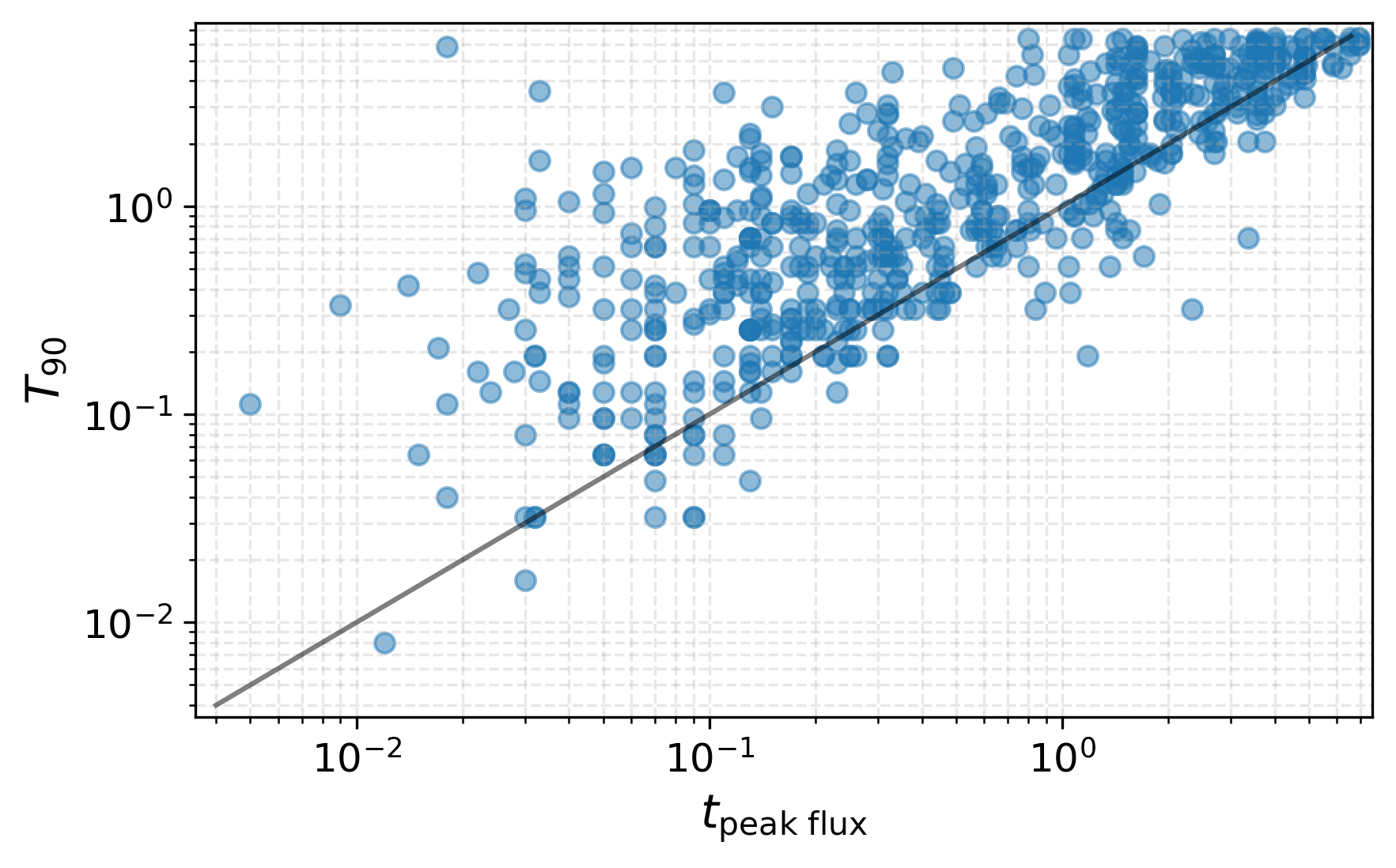}
                \includegraphics[width=\columnwidth]{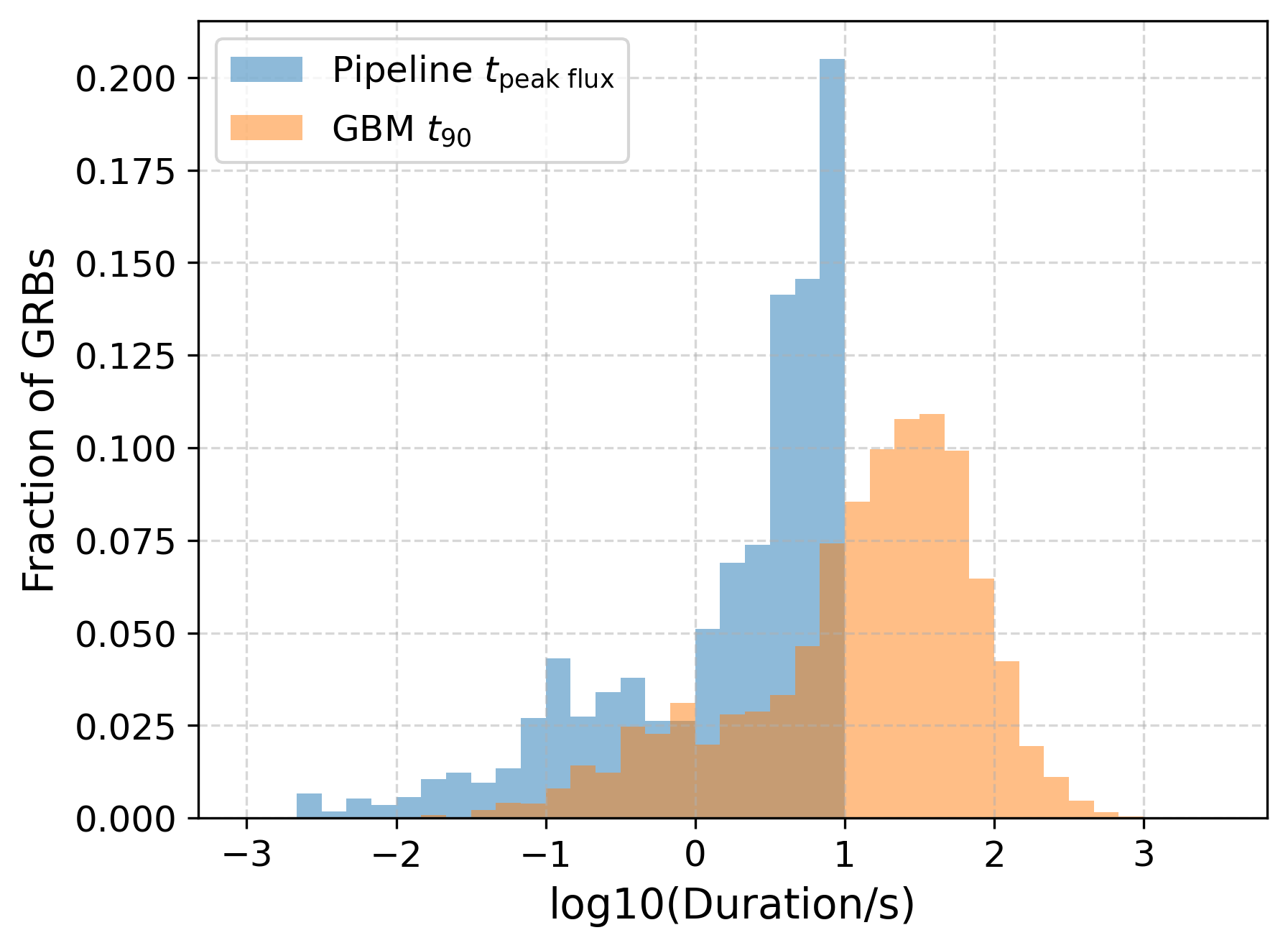}
                \caption{\textbf{Duration comparison between the pipeline and all GRBs from the GBM catalog.}
                \textit{Top:} Comparison of the pipeline peak-flux duration (defined from the best-fit box template) and the GBM catalog $T_{90}$ for bursts common to both samples, restricted to $T_{90} < 6.57\,\mathrm{s}$. The straight line indicates equality. Above the line $T_{90}$ exceeds the pipeline duration. The plot shows that the peak-flux durations are systematically shorter than the catalog $T_{90}$ values for most bursts.
                \textit{Bottom:} Normalized duration distributions. The blue histogram shows $\log_{10}(\mathrm{duration/s})$ for GRB-classified pipeline events with $p_{\mathrm{astro}} \geq 0.9$, while the orange histogram shows the GBM catalog $T_{90}$ distribution. The pipeline durations are shifted toward shorter times compared to the GBM sample.}
                \label{fig: duration comp}
            \end{figure}

            Directly comparing duration parameters requires care, as the definitions differ between catalogs. The GBM catalog reports $T_{90}$ (the interval accumulating 90\% of the fluence), whereas our pipeline reports the duration of the boxcar template maximizing the likelihood (approximating the duration of the peak flux emission). The top panel of Fig. \ref{fig: duration comp} shows a scatter plot of the $T_{90}$ duration from the GBM catalog and our estimated peak flux duration for GRBs jointly detected by our pipeline and the GBM catalog. The peak-flux duration estimates are systematically shorter than the corresponding full burst durations. 
            
            In addition, a comparison between the duration distribution of our GRB triggers and that of the GBM catalog (lower panel of Fig. \ref{fig: duration comp}) shows overall consistency, once the systematic duration bias discussed above and the upper duration limit of our search are taken into account. The bimodality remains visible, although the separation occurs at shorter durations in our sample. Since our search is restricted to bursts with durations of approximately 6 seconds, an accumulation of triggers is observed near the maximum duration permitted by our peak-flux duration estimator (see Sec. \ref{sec: parameter estimation}).
            
            Very short peak-flux episodes identified by our pipeline may, in some cases, correspond to the brightest emission episodes of much longer GRBs. This behavior is expected, as the pipeline was specifically optimized for short-duration events, which can lead to portions of long GRBs being absorbed into the estimated background. In a sense, this is similar to a bias known as the "tip-of-the-iceberg" effect that has been discussed in the literature as a source of contamination between the long and short GRB populations (see, e.g., \cite{moss_instrumental_2022}). We find that approximately 6\% of the GRBs in our catalog with peak-flux durations shorter than 2 seconds are associated with GBM catalog events whose reported $T_{90}$ durations are more than two orders of magnitude longer. Restricting the sample to peak-flux durations shorter than 0.1 seconds, approximately 3\% are associated with GBM events whose $T_{90}$ durations are more than three orders of magnitude longer. These results suggest that caution should be taken when interpreting the shortest-duration triggers.

        \subsubsection{Spectral parameter consistency}
        To assess the consistency of the spectrum of our GRB triggers, we compare the MLE of the Band function parameters ($\alpha$, $\beta$, and $E_{\text{peak}}$) against the GBM catalog values.
        
        \begin{figure*}
            \centering
            \includegraphics[width=\textwidth]{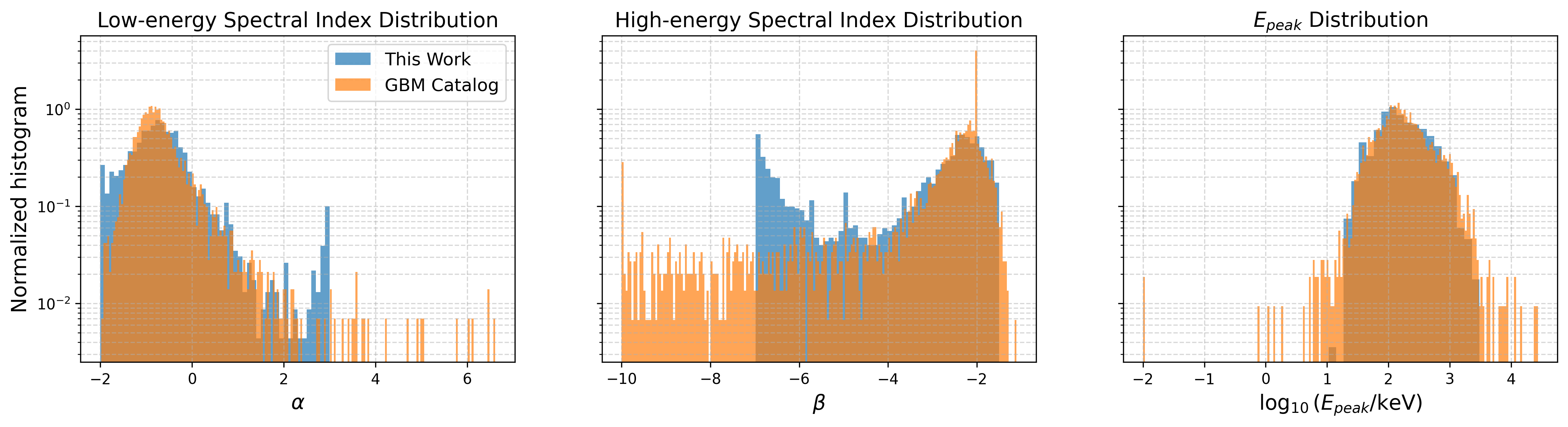}
            \caption{\textbf{Spectral parameter distributions.} We compare the MLE spectral parameters from our pipeline (blue) with the GBM catalog values (orange). \textit{Left:} The low-energy spectral index $\alpha$. \textit{Center:} The high-energy spectral index $\beta$. \textit{Right:} The peak energy $E_{\text{peak}}$. Our recovered parameters largely track the catalog, with deviations in $\alpha$ and $\beta$ arising from our cutoff values imposed when sampling the likelihood (Eq. \ref{eq: posterior}).}
            \label{fig: spectral comp}
        \end{figure*}
        
        As shown in Figure \ref{fig: spectral comp}, the distributions of the low-energy index $\alpha$ and the peak energy $E_{\text{peak}}$ are in good agreement with the known population. The distribution of the spectral indices $\alpha$ and $\beta$ shows sharp peaks in the histograms corresponding to this work (e.g., near $\alpha \approx 3$ and $\beta \approx -6$) as a result of the uniform prior limits we set to the sampler.
        
    \subsection{Comparison with the Swift BAT} \label{sec: BAT comp}
        \subsubsection{BAT joint detection analysis}
            As discussed in Sec \ref{sec: bat followup}, we performed a targeted search in Swift/BAT rates data to find if we have a joint detection. Here, a jointly detected event refers to a GBM trigger with a statistically significant counterpart in the Swift/BAT rate data ($p_{\text{bat}}\geq0.9$). Our analysis reveals that with $p_{\text{bat}}\geq0.9$ we have a total of 1736 jointly detected events, out of which 1277 are new events (that is, not recorded in the GBM catalog). Restricting to the GRB classes alone, there are 567 jointly detected events, out of which 192 are not in the GBM catalog. Among these 192 events, 21 are associated with GRBs listed in the Swift/BAT GRB catalog\footnote{\url{https://swift.gsfc.nasa.gov/results/batgrbcat/}}. All 21 have high $p_{\text{bat}}$ values, indicating highly significant detections in the BAT data, and their estimated sky positions are consistent with the reported BAT localizations. Their properties are listed in Table \ref{tab: bat no gbm} in Appendix \ref{appendix: tables}. We also searched the GCN archive and found that only one of the events was detected by the GBM team subthreshold search, while four others had previously been identified through targeted GBM team follow-up analyses of BAT-triggered GRBs. These results demonstrate that our joint detection analysis can substantially increase the significance of marginal detections, enabling the recovery of astrophysical events that would otherwise remain buried in the noise.
            
            The presence of an independent detection by Swift/BAT provides an external validation of our pipeline, reducing the likelihood of instrumental or analysis artifacts and increasing confidence in the astrophysical origin of these events. In particular, the recovery of a substantial number of previously uncataloged GRBs in both instruments demonstrates the validity of our search methodology.

        \subsubsection{BAT GRB catalog}
    
            \begin{figure*}
            \centering
                \includegraphics[scale=0.55]{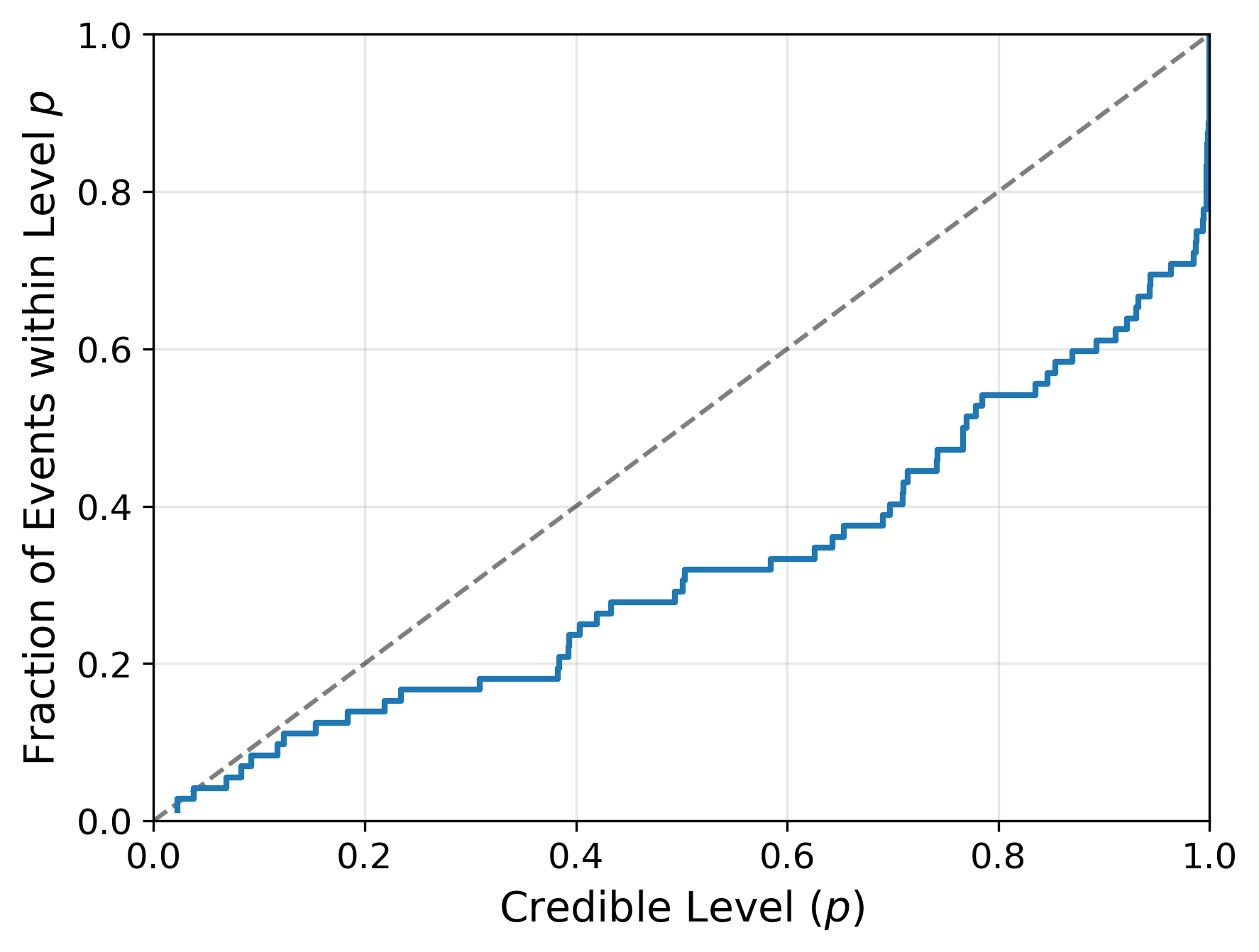}
                \includegraphics[scale=0.55]{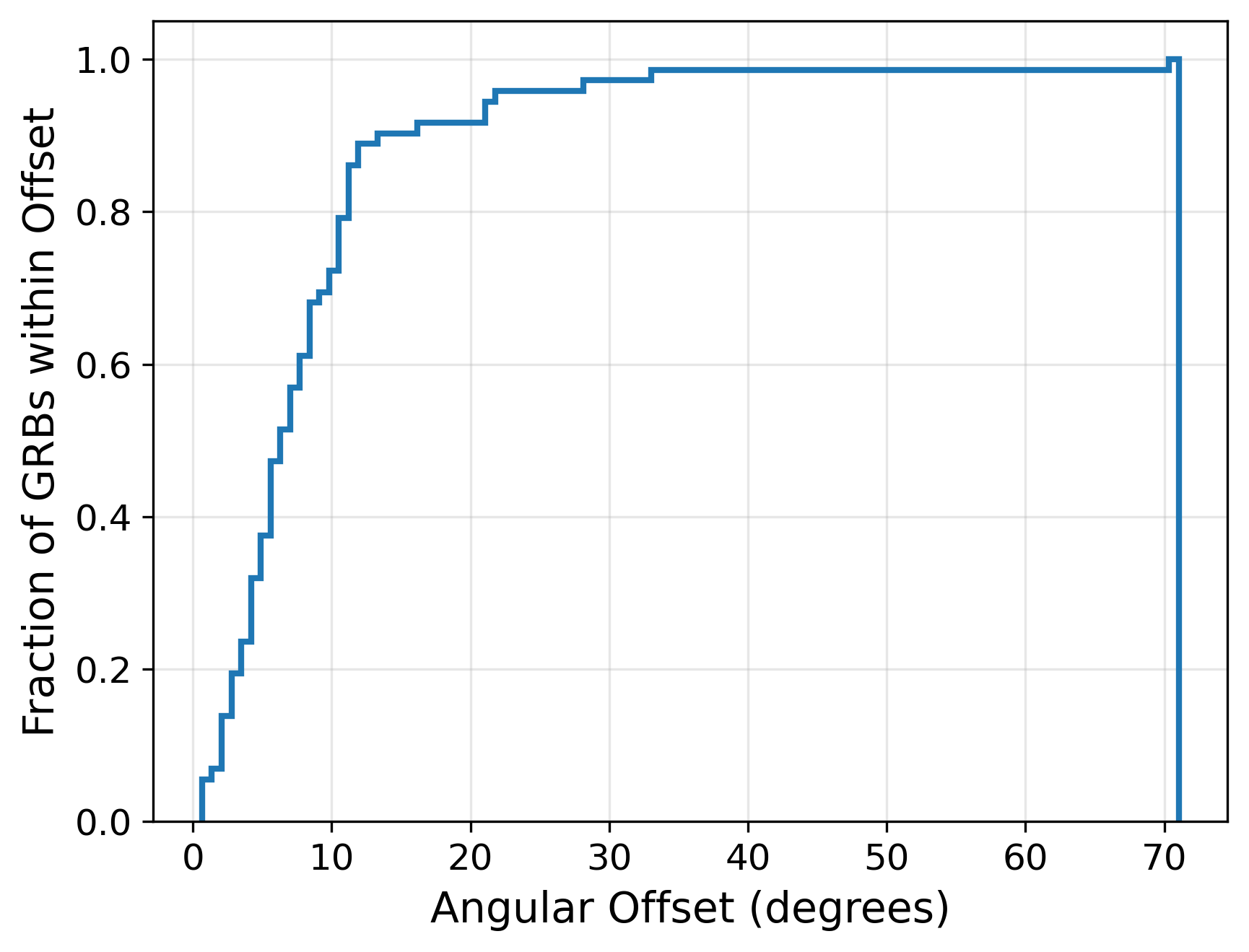}
                \caption{\textbf{Systematic assessment of our localization performance}. 
                We use the \textit{Swift}/BAT positions of jointly detected GRBs as the reference sky locations to evaluate our localizations. 
                \textit{Left:} P-P plot showing the fraction of events contained within a given highest posterior density credible level, computed by integrating the posterior probability from the maximum down to the posterior density at the corresponding Swift/BAT position. The deviation from the diagonal expectation demonstrates that the reported credible regions are systematically underestimated.
                \textit{Right:} Cumulative distribution of angular offsets between our maximum-likelihood positions and the corresponding BAT locations. The distribution indicates a systematic localization offset.} 
            \label{fig: localization tests}
            \end{figure*}
            
            \begin{figure*}
                \centering
                \includegraphics[scale=0.6]{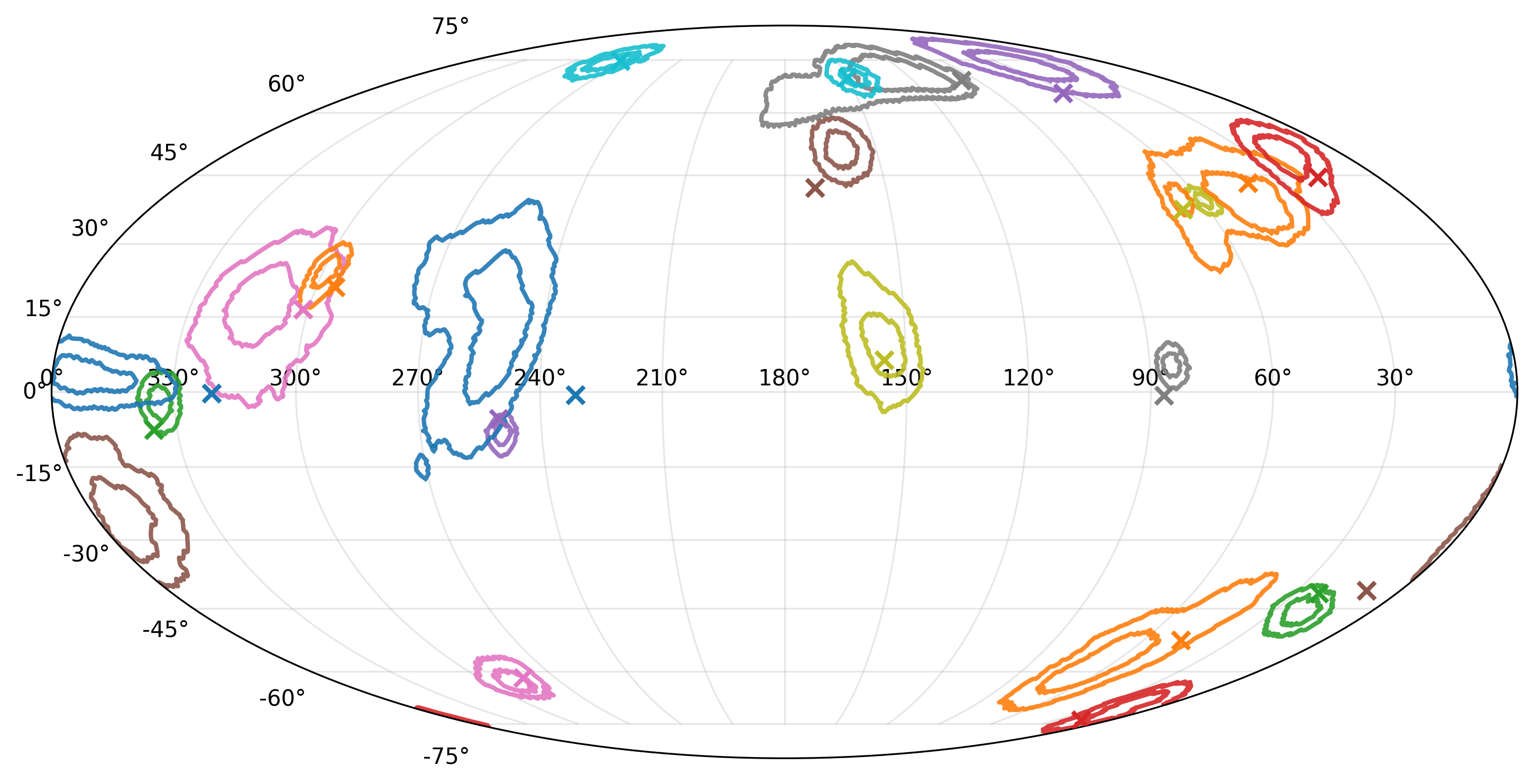}
                \caption{\textbf{Localization contours for a sample of GRBs jointly detected by our pipeline and Swift/BAT.} Solid lines represent the 50\% and 90\% credible regions derived from our parameter estimation framework, while crosses mark the reference Swift/BAT sky positions. To ensure visual clarity and prevent overlapping contours, this display is restricted to events occurring from 2020 onward, and the sample is limited to GRBs with durations shorter than our maximum search template.}
                \label{fig: localization contours}
            \end{figure*}
            
            We use the Swift/BAT GRB catalog updated through 2025-06-05, to provide an independent comparison between our pipeline detections and an external catalog, as well as to assess our localization performance. Since BAT is more sensitive than GBM and provides substantially more accurate localizations, we treat BAT as a reference (“ground truth”) for this cross-catalog comparison.
            
            During our search period (2013–2025), the BAT catalog contains a total of 163 GRBs. For each BAT GRB, we search for a corresponding detection within a $\pm 20$ s window in both our pipeline output and the GBM catalog. When enforcing a duration consistency requirement by restricting matches to $T_{90} \le 6.57$ s, corresponding to the longest duration allowed by our estimator, we find 72 BAT GRBs with matches in our pipeline and 75 with matches in GBM. We further identify 7 BAT GRBs matched by our pipeline but not by GBM, and 10 BAT GRBs matched by GBM but not by our pipeline. Of these 10 missed triggers, 3 were skipped in the pre-processing stage (for two, half of the detectors were inactive, and the other trigger happened at the edge of a TTE file), and 7 were rejected by our pipeline's vetoes (6 resulting from very high SNR characteristics, and 1 due to low SNR). These are summarized in Table \ref{tab: missed bat} of Appendix \ref{appendix: tables}.
            
            If the duration restriction is relaxed, allowing long BAT GRBs to be associated with our short-spike detections, the number of cross-matches increases. However, more long-duration GRBs remain undetected by our pipeline, as it is not calibrated for such events. In particular, some of the processing steps we apply, such as the background drift correction, are known to suppress long-duration GRBs by interpreting their slowly varying count rates as background variations. The results are summarized in Table \ref{tab: bat comparison}.
    \begin{table}
    \centering
    \caption{Comparison with the \textit{Swift}/BAT GRB Catalog (2013--2025).}
    \label{tab: bat comparison}
    \begin{tabular}{lcc}
        \hline
         & \multicolumn{2}{c}{Number of Events} \\
        \hline
        Total \textit{Swift}/BAT Events & \multicolumn{2}{c}{163} \\
        \textit{Swift}/BAT Events Visible to GBM & \multicolumn{2}{c}{105} \\
        \hline
        & $p_{\text{astro}} \geq 0.9$ & $p_{\text{astro}} \geq 0.5$ \\
        \hline
        Matches in Our Pipeline         & 72      & 76 \\
        Matches in GBM Catalog          & 75       & 75 \\
        \hline
        Our Pipeline Only               & 7 & 11 \\
        GBM Only                        & 10  & 10 \\
        \hline
    \end{tabular}
    \medskip\\
    {\footnotesize \textit{Note.} Comparison of our pipeline's recovery against a reference set of 163 \textit{Swift}/BAT GRBs ($T_{90} < 6.573$\,s) detected between 2013 and 2025. Events "Visible to GBM" ($N=105$) exclude those occurring while \textit{Fermi} was in the South Atlantic Anomaly (SAA) or when the burst position was Earth-occulted for GBM. Pipeline performance is evaluated at two astrophysical probability thresholds ($p_{\text{astro}}$) and includes only triggers in classes GRB1 through GRB7.}
    \end{table}

            We use the jointly detected GRBs, listed together with their relevant properties in Table \ref{tab:contour_sample} in Appendix \ref{appendix: tables}, to evaluate the localization performance of our pipeline's parameter estimation.
            We do this by constructing the P-P (percentile-percentile) plot, which is the cumulative fraction of events falling inside each credible level. Here, the credible level is the highest posterior density credible level computed by integrating the posterior probability from its maximum down to the posterior density at the reference Swift/BAT position. A perfectly calibrated localization will result in an $x$ fraction of events falling with $x$ confidence level - i.e., on the straight line in the P-P plot. Also, we calculate the angular distance between the BAT localization and our MLE of the position. The results are presented in Fig. \ref{fig: localization tests}. The systematic is clearly visible as the fraction of events within the 90\% credible level is $\sim 70\%$. In addition, $\sim 70\%$ of our localizations are within $\sim 10^{\circ}$. To provide a visual representation of this performance, Fig. \ref{fig: localization contours} displays the 50\% and 90\% confidence contours derived from our parameter estimation alongside the reference Swift/BAT positions for a subset of jointly detected GRBs (restricted to post-2020 events to avoid overcrowding). This sky map qualitatively illustrates the systematic offsets quantified in Fig. \ref{fig: localization tests}, showing how the precise BAT localizations frequently fall near the edges of, or just outside, our nominal 90\% credible regions. It should be emphasized again that the PE framework we apply was designed to trade off accuracy and computation time. An improved localization can be achieved by generating accurate responses at the time of the event, increasing the number of samples, and choosing between multiple models and not sticking to the Band function alone.

\section{Summary and conclusions} \label{sec: summary}
    In this paper, we presented the application of our GRB detection pipeline to the complete GBM CTTE data set spanning 2013–2025. In doing so, we developed a new framework for parameter estimation and trigger classification based on the global structures revealed by the large amounts of bursts and their inferred parameters. A systematic comparison between our localizations and those reported by \textit{Swift}/BAT shows agreement within the known GBM systematic uncertainties. While further refinement is possible, our localization approach is computationally efficient and avoids generating burst-specific response matrices, which would have rendered a full archival search impractical.
    
    Compared to previous GBM subthreshold searches, the framework presented here provides improved sensitivity through a fully coherent Poisson matched-filter search using extensive spectral and temporal template banks, a statistically calibrated probability derived from timeslides for each candidate, an automated classification framework for the heterogeneous transient population, and a reproducible, open-source implementation.
    
    A central outcome of this work is the probabilistic interpretation assigned to every trigger. By applying the same analysis to timeslided data, which contain no true astrophysical transients, we constructed an empirical background distribution and assigned each event a probability of astrophysical origin, $p_{\text{astro}}$, along with a corresponding false-alarm rate. This quantitative measure substantially expands the set of bursts that can be meaningfully compared with external transient catalogs, including gravitational waves from the LIGO-Virgo-KAGRA (LVK) collaboration and fast radio bursts (FRBs) detected by CHIME/FRB (\cite{chimefrb_collaboration_chime_2018}).
    
    We also performed a systematic followup search with \textit{Swift}/BAT. Using the same statistical framework underlying $p_{\text{astro}}$, we defined $p_{\text{bat}}$ as the probability that a signal is detected in Swift BAT rather than a statistical fluctuation. Joint detections strengthen the credibility of marginal events and provide sensitivity to fainter transients that may fall below the standalone detection threshold of either instrument. Furthermore, the methodology developed here naturally extends beyond mere validation. Because the combined information from both instruments can yield a statistically significant trigger even when each instrument alone would not, this approach can be used in future joint searches to uncover new populations of low-SNR GRBs operating below current detection thresholds. A detailed presentation of this methodology and its broader applications will be discussed in a companion paper.
    
    Our broad search, unconstrained by specific detectors or energy selections, uncovered a large population of SGR bursts. These were classified within our new framework, and each burst is assigned both $p_{\text{astro}}$ and $p_{\text{bat}}$. Combined with temporal clustering analyses and BAT localization maps when available, these probability measures can aid in future source associations. The resulting SGR catalog, together with BAT localization information, may help characterize known Magnetars and identify previously unknown sources. Their properties and associations will be examined in a dedicated follow-up study.
    
    The large number of probabilistically ranked triggers also opens the possibility of identifying rare or previously unrecognized transient populations. As an example, our pipeline is sensitive to millisecond-scale peak-flux episodes. While some of these may correspond to ultra-short transients, others may represent the brightest emission episodes of longer, distant GRBs. Distinguishing between these possibilities requires careful event-by-event analysis. Nevertheless, blind searches have not systematically explored this millisecond peak-flux regime, which may contain phenomena deserving further follow-up.
    
    Finally, while the present pipeline is not optimized for long-duration bursts, the statistical framework introduced here can be extended to matched-filter strategies tailored to longer timescales. Such developments could further enrich GRB and SGR studies and, more broadly, benefit the hard X-ray and gamma-ray transient community.

\section*{Data availability}
    The detection pipeline code and parameter estimation framework are publicly available on GitHub at \url{https://github.com/PeAriel/grpype}.
    The trigger list presented in this work, filtered to $p_{\text{astro}}\geq0.5$ or $p_{\text{bat}}\geq0.9$, is publicly available on Zenodo \citep{perera_catalog_2026}\footnote{\url{https://doi.org/10.5281/zenodo.20430377}} as the file \texttt{perera\_et\_al\_2026\_gbm\_grb\_catalog.csv}. Posterior samples for most of the triggers in this list with SNR${}^2\geq 50$ are available in the same Zenodo repository as the file \texttt{perera\_et\_al\_2026\_posterior\_samples.tar.gz}.
    Additional data products generated during this study are available from the corresponding author upon reasonable request.
    
\section*{Acknowledgments}
    We thank Jimmy Delaunay for valuable information and insightful discussions about the BAT instrument and data products.
    This research was supported by grant no 2022136 from the United States - Israel Binational Science Foundation
    (BSF), Jerusalem, Israel.
    BZ is supported by a research grant from the Willner Family Leadership Institute for the Weizmann Institute of Science. 
    TV additionally acknowledges support from NSF grants 2012086 and 2309360, the Alfred P. Sloan Foundation through grant number FG-2023-20470, and the Hellman Family Faculty Fellowship during the time this work was done.


\bibliographystyle{mnras}
\bibliography{references}



\appendix

\section{Derivation of the BAT follow-up test statistic} \label{appendix: BAT statistic}
    In this appendix, we derive the test statistic used for the Swift/BAT follow-up analysis. To get the optimal test in the presence of Gaussian noise and without free model parameters, we use the likelihood ratio test.
    
    Let $d_t$ represent the observed number of counts rate from the BAT instrument rates data. We denote the null hypothesis as $\mathcal{H}_0$ (the noise only hypothesis) and the alternative hypothesis as $\mathcal{H}_1$ (the presence of a GRB signal). The BAT rates are the number of photons per unit time, integrated over the detectors and their energy channels, and hence contain sufficient background counts to be modeled as Gaussian.
    
    The two hypotheses are
    \begin{equation}
        \begin{aligned}
           &\mathcal{H}_0: \; d_{t} \sim \mathcal{N}(b_t, \sigma_t^2); \\
           &\mathcal{H}_1: \; d_{t} \sim \mathcal{N}(b_t + AT_{t_0}(\Delta), \sigma_t^2),
        \end{aligned}
    \end{equation}
    Where $b_t$ is the mean background count rate at time $t$, $\sigma_t$ is the standard deviation of the background count rate at time $t$, $T_{t_0}(\Delta)$ is the signal template at time $t_0$ with duration $\Delta$, and $A$ is the signal amplitude. Since the burst shape is unknown, we use a boxcar signal with the width $\Delta$ of the GRB as determined by our GBM detection pipeline, placed symmetrically around the trigger time $t_0$. Since there is only one free parameter in this model, the amplitude becomes irrelevant as we show in the likelihood ratio below.
    
    The mean background count rate and its standard deviation are unknown and time variable. For these reasons, we estimate them using a rolling mean with a gap as described on PZV25 Sec 3.3.2, we use $\hat{b}_t$ and $\hat{\sigma}_t$ instead of $b_t$ and $\sigma_t$, respectively, in the calculation of the test statistic.

    With all these components, the likelihood ratio test is
    \begin{equation}
        L = \frac{P(d_t|\mathcal{H}_1)}{P(d_t|\mathcal{H}_0)} = \prod_{t} \exp\left( \frac{(d_t - \hat{b}_t)^2 - (d_t - \hat{b}_t - AT_{t_0}(\Delta))^2}{2\hat{\sigma}_t^2} \right).
    \end{equation}
    As the application of any monotonic function does not change the test but only the detection threshold (which we determine for the final test form), we can take the logarithm of the likelihood ratio to get the test statistic
    \begin{equation}
        \log L = \sum_{t} \left( \frac{(d_t - \hat{b}_t)^2 - (d_t - \hat{b}_t - AT_{t_0}(\Delta))^2}{2\hat{\sigma}_t^2} \right).
    \end{equation}
    Simplifying the expression, and noting that any term independent of $d_t$ does not affect the test statistic and can be ignored (e.g., by the application of a monotonic function), we get
    \begin{equation}
        \log L = \sum_{t} \frac{A(d_t - \hat{b}_t)T_{t_0}(\Delta)}{\hat{\sigma}_t^2}.
    \end{equation}
    To get a more illuminating form, we can apply a monotonic function to the test statistic. In particular, we can normalize the test statistic to have 0 expected value and unit variance under the null hypothesis. This is achieved by defining the test statistic as
    \begin{equation}
        E[\log L | \mathcal{H}_0] = 0, \quad \text{Var}[\log L | \mathcal{H}_0] = \sum_{t} \frac{A^2 T_{t_0}(\Delta)^2}{\hat{\sigma}_t^2}.
    \end{equation}
    Defining the test statistic $\mathcal{S}$ as this normalized version, we get
    \begin{equation}
        \mathcal{S} = \frac{\log L - E[\log L | \mathcal{H}_0]}{\sqrt{\text{Var}[\log L | \mathcal{H}_0]}} = \sum_{t}\frac{(d_t - \hat{b}_t)T_{t_0}(\Delta)}{\hat{\sigma}_t^2\sqrt{\sum_{t} \frac{T_{t_0}^2(\Delta)}{\hat{\sigma}_t^2}}}.
    \end{equation}
    As a last step, we assume that the background variance is approximately constant over the duration of the burst, i.e., $\hat{\sigma}_t \approx \hat{\sigma}_{t_0}$, which simplifies the test statistic to
    \begin{equation}
        \mathcal{S} = \sum_{t=t_0-\Delta/2}^{t_0+\Delta/2}\frac{(d_t - \hat{b}_t)}{\hat{\sigma}_{t_0} \sqrt{\Delta}}.
    \end{equation}
    This is the test statistic we use for the BAT follow-up analysis as described in Sec. \ref{sec: bat followup}, Eq. \ref{eq: BAT statistic}.

\section{Trigger Catalog} \label{sec:appendix_catalog}

The complete 13-year trigger catalog and posterior samples for events with SNR${}^2\geq50$ are available in Zenodo \citep{perera_catalog_2026}. A small portion of the trigger catalog is shown in Table \ref{tab: catalog_stub} for reference. Descriptions of the catalog columns are provided below.

\begin{itemize}
    \item \textbf{\texttt{trigtime}}: The trigger time of the transient in UTC.
    \item \textbf{\texttt{duration}}: The duration of the search boxcar template (in seconds) that maximized the detection test statistic.
    \item \textbf{\texttt{search\_binning}}: The time resolution of the TTE data segment used during the initial search pipeline (either 1 ms or 10 ms).
    \item \textbf{\texttt{slc\_index}}: The internal index of the data slice analyzed by the pipeline. This information was added for the reproducibility of the results. Relevant only for \texttt{search\_binning==0.001} sec.
    \item \textbf{\texttt{pe\_duration}}: The refined peak-flux duration (in seconds) derived during the parameter estimation stage.
    \item \textbf{\texttt{snr}}: The calibrated, drift-corrected Poisson matched-filter signal-to-noise ratio from the detection pipeline. Reported in units of standard normal distribution $\sigma$.
    \item \textbf{\texttt{pastro}}: The probability of astrophysical origin ($p_{\rm astro}$).
    \item \textbf{\texttt{FAR}}: The false alarm rate of the trigger, in units of Hz. Calculated assuming 65\% livetime of GBM over 13 years.
    \item \textbf{\texttt{bat\_snr}}: The test statistic ($\rho_{\rm bat}$) resulting from the Swift/BAT targeted follow-up search. Reported in units of standard normal distribution $\sigma$.
    \item \textbf{\texttt{pbat}}: The probability that a temporally coincident signal in Swift/BAT is a genuine joint detection ($p_{\rm bat}$).
    \item \textbf{\texttt{classification}}: The source class assigned to the trigger by our decision tree framework (e.g., GRB1).
    \item \textbf{\texttt{gbm\_catalog}}: The identifier of a temporally coincident event in the GBM catalog.
    \item \textbf{\texttt{ra\_median}, \texttt{dec\_median}}: The median values of the marginalized posterior distribution for the Right Ascension and Declination (in degrees).
    \item \textbf{\texttt{ra\_err\_plus}, \texttt{dec\_err\_plus}}: The positive 1$\sigma$ error bounds for the median values of the marginalized posterior distribution for the Right Ascension and Declination (in degrees).
    \item \textbf{\texttt{ra\_err\_minus}, \texttt{dec\_err\_minus}}: The negative 1$\sigma$ error bounds for the median values of the marginalized posterior distribution for the Right Ascension and Declination (in degrees).
    \item \textbf{\texttt{band\_epeak\_median}, \texttt{band\_alpha\_median}, \texttt{band\_beta\_median}}: The median values of the marginalized posterior for the Band function spectral parameters ($E_{\rm peak}$ in keV, $\alpha$, and $\beta$).
    \item \textbf{\texttt{band\_epeak\_err\_plus}, \texttt{band\_alpha\_err\_plus}, \texttt{band\_beta\_err\_plus}}: The positive 1$\sigma$ error bounds for the median values of the marginalized posterior for the Band function spectral parameters ($E_{\rm peak}$ in keV, $\alpha$, and $\beta$).
    \item \textbf{\texttt{band\_epeak\_err\_minus}, \texttt{band\_alpha\_err\_minus}, \texttt{band\_beta\_err\_minus}}: The negative 1$\sigma$ error bounds for the median values of the marginalized posterior for the Band function spectral parameters ($E_{\rm peak}$ in keV, $\alpha$, and $\beta$).
    \item \textbf{\texttt{ra\_max}, \texttt{dec\_max}}: The Maximum Likelihood Estimates (MLE) for the sky position.
    \item \textbf{\texttt{band\_epeak\_max}, \texttt{band\_alpha\_max}, \texttt{band\_beta\_max}}: The Maximum Likelihood Estimates (MLE) for the Band function spectral parameters.
    \item \textbf{\texttt{earth\_bayes\_factor}}: A statistical measure quantifying the probability of the trigger originating from the Earth.
    \item \textbf{\texttt{sun\_bayes\_factor}}: A statistical measure quantifying the probability of the trigger originating from the Sun.
\end{itemize}

\section{Tables} \label{appendix: tables}

\begin{table*}
\centering
\caption{Pipeline triggers detected in the BAT rates data, are present in the Swift/BAT GRB catalog and are not included in the GBM catalog.}
\label{tab: bat no gbm}
\begin{tabular}{llccccccccc}
\hline
GRB Name & Trigger Time & BAT $T_{90}$ & $p_{\rm astro}$ & $p_{\rm bat}$ & BAT RA & BAT Dec & Pipe RA & Pipe Dec & Offset & GBM subthresh. \\
 & (UTC) & (s) &  &  & (deg) & (deg) & (deg) & (deg) & (deg) &  GCN Circ. \\
\hline
GRB131117A & 2013-11-17 00:34:08.917 & 10.88 & 0.759 & 1.000 & 332.35 & -31.76 & 356.45 & -27.80 & 21.24 & -- \\
GRB131224B & 2013-12-24 03:25:09.136 & 8.35 & 1.000 & 0.996 & 163.72 & -14.18 & 159.40 & -18.06 & 5.68 & -- \\
GRB140301A & 2014-03-01 15:24:50.987 & 27.80 & 0.532 & 0.979 & 69.53 & -34.24 & 60.00 & -34.40 & 7.86 & -- \\
GRB140515A & 2014-05-15 09:12:34.216 & 23.42 & 0.000 & 0.996 & 186.07 & 15.10 & 181.48 & 21.04 & 7.36 & -- \\
GRB140516A & 2014-05-16 20:30:54.846 & 0.26 & 0.408 & 0.983 & 252.98 & 39.96 & 259.95 & 28.91 & 12.45 & -- \\
GRB150202A & 2015-02-02 23:10:04.873 & 25.47 & 0.579 & 0.999 & 39.24 & -33.13 & 18.90 & -12.81 & 27.52 & -- \\
GRB151029A & 2015-10-29 07:49:40.224 & 8.95 & 1.000 & 1.000 & 38.54 & -35.36 & 33.72 & -30.80 & 6.09 & -- \\
GRB151114A & 2015-11-14 09:59:34.553 & 4.86 & 0.511 & 1.000 & 120.95 & -61.03 & 115.35 & -68.37 & 7.71 & -- \\
GRB160327A & 2016-03-27 09:16:12.043 & 33.74 & 1.000 & 1.000 & 146.70 & 54.02 & 163.48 & 61.46 & 11.58 & -- \\
GRB160417A & 2016-04-17 04:23:44.337 & 14.55 & 0.479 & 0.996 & 120.26 & 7.65 & 117.48 & -0.94 & 9.03 & -- \\
GRB170516A & 2017-05-16 12:49:22.501 & 36.77 & 0.477 & 0.983 & 41.54 & -55.93 & 36.24 & -54.37 & 3.41 & -- \\
GRB170728A & 2017-07-28 06:53:29.621 & 1.25 & 0.576 & 0.983 & 58.90 & 12.16 & 55.32 & 23.33 & 11.67 & -- \\
GRB200409A & 2020-04-09 03:20:00.205 & 16.00 & 0.999 & 1.000 & 60.56 & -50.24 & 53.47 & -43.09 & 8.64 & 27514 \\
GRB200729A & 2020-07-29 19:38:10.468 & 122.00 & 0.773 & 0.983 & 184.36 & 45.57 & 190.67 & 39.60 & 7.56 & 28174 \\
GRB210104B & 2021-01-04 21:10:21.215 & 19.94 & 0.468 & 1.000 & 53.56 & 37.95 & 53.16 & 36.63 & 1.36 & -- \\
GRB210222B & 2021-02-22 22:37:25.718 & 12.11 & 0.999 & 0.996 & 154.63 & -14.94 & 160.57 & -8.67 & 8.55 & -- \\
GRB210725B & 2021-07-25 12:00:50.213 & 417.91 & 0.057 & 0.996 & 192.89 & 17.10 & 186.96 & 8.97 & 9.97 & 30546 \\
GRB210905A & 2021-09-05 00:12:40.085 & 778.40 & 0.633 & 1.000 & 309.05 & -44.42 & 327.26 & -38.50 & 14.83 & 30779 \\
GRB210930A & 2021-09-30 02:47:01.752 & 11.81 & 1.000 & 1.000 & 197.41 & 48.63 & 184.75 & 44.32 & 9.71 & -- \\
GRB211207A & 2021-12-07 20:52:57.951 & 3.73 & 0.776 & 0.983 & 149.60 & -24.36 & 156.98 & -11.39 & 14.74 & 31290 \\
GRB230216A & 2023-02-16 14:48:35.530 & 91.20 & 0.999 & 0.983 & 113.97 & -7.99 & 120.60 & -18.42 & 12.26 & -- \\
\hline
\end{tabular}
\medskip\\
{\footnotesize \textit{Note.} 
The table lists the 21 Swift/BAT GRBs that were found by our BAT rates follow-up search and are not present in the GBM catalog. Column 1 gives the BAT GRB name, followed by the pipeline trigger time (UTC), the BAT $T_{90}$ duration, the astrophysical probability ($p_{\rm astro}$), and the significance of the BAT counterpart ($p_{\rm BAT}$). Columns 6-9 list the reported BAT localization and the localization estimated by our pipeline, while Column 10 gives the angular separation between the two positions. The final column lists the corresponding GCN Circular number when a targeted GBM follow-up analysis of the BAT trigger was reported.}
\end{table*}

\begin{table*}
\centering
\caption{Trigger Catalog Table Example.}
\label{tab: catalog_stub}
\begin{tabular}{lcccccccccc}
    \hline
    \texttt{trigtime} & 
    \texttt{duration} & 
    \texttt{snr} & 
    \texttt{pastro} & 
    \texttt{bat\_snr} & 
    \texttt{pbat} & 
    \texttt{classification} & 
    \texttt{ra\_max} & 
    \texttt{dec\_max} &
    \texttt{band\_epeak\_max} &
    $\cdots$ \\
    (UTC) & 
    (s) & 
    ($\sigma$) & 
    & 
    ($\sigma$) & 
    & 
    & 
    (deg) & 
    (deg) &
    (keV) &
    \\
    \hline
2013-01-01 17:15:58.278 & 4.869 & 9.020 & 0.788 & $-$0.239 & 0.007 & GRB5 & 108.132 & $-$33.387 & 83.198  & $\cdots$ \\
2013-01-02 00:05:04.712 & 4.869 & 8.806 & 0.502 & 2.516 & 0.302 & GRB3 & 120.895 & $-$48.670 & 190.829  & $\cdots$ \\
2013-01-02 19:40:56.050 & 0.003 & 19.573 & 1.000 & 0.390 & 0.020 & (GRB/TGF)2 & 354.629 & 9.779 & 2999.769  & $\cdots$ \\
2013-01-02 20:39:23.259 & 6.573 & 7.926 & 0.529 & 2.547 & 0.305 & GRB1 & 133.674 & 22.634 & 34.621  & $\cdots$ \\
2013-01-04 05:13:35.063 & 6.573 & 8.628 & 0.688 & 0.561 & 0.041 & GRB5 & 287.297 & $-$17.150 & 32.981  & $\cdots$ \\
$\cdots$ & $\cdots$ & $\cdots$ & $\cdots$ & $\cdots$ & $\cdots$ & $\cdots$ & $\cdots$ & $\cdots$ & $\cdots$ & $\cdots$ \\
    \hline
\end{tabular}
\medskip\\
{\footnotesize \textit{Note.} This table is published in its entirety on Zenodo \citep{perera_catalog_2026}. A portion is shown here for guidance regarding its form and content.}
\end{table*}

\begin{table}
\centering
\caption{Summary of \textit{Swift}/BAT events present in the GBM catalog and missed by our pipeline.}
\label{tab: missed bat}
\begin{tabular}{cl}
    \hline
    Trigger Time (UTC) & Reason for Exclusion \\
    \hline
2013-02-16 22:15:24.622 & High SNR; rejected by occultation veto \\
2016-07-09 19:49:03.464 & High SNR; rejected by particle precipitation veto \\
2018-02-04 02:36:16.941 & High SNR; rejected by particle precipitation veto \\
2019-10-31 21:23:31.132 & High SNR; rejected by particle precipitation veto \\
2021-03-23 22:02:18.464 & High SNR; rejected by particle precipitation veto \\
2021-06-18 01:43:37.008 & Marginal SNR ($\approx 7$); timeslides at the trigger time had higher SNR \\
2023-02-17 21:53:10.903 & Several inactive detectors \\
2024-01-12 17:37:24.977 & High SNR; rejected by particle precipitation veto \\
2024-02-22 08:51:31.328 & Several inactive detectors \\
2024-06-24 06:01:09.993 & Data processing exclusion (unprocessed file boundary) \\
    \hline
\end{tabular}
\medskip\\
{\footnotesize \textit{Note.} A detailed breakdown of the 10 \textit{Swift}/BAT jointly detected GRBs that were recovered by the GBM catalog but missed or rejected by our automated pipeline.}
\end{table}

\onecolumn
\begin{longtable}{llcccccccc}
\caption{Sample of Jointly Detected GRBs.} \label{tab:contour_sample} \\
\hline
GRB Name & Pipeline Trigger Time & BAT RA & BAT Dec & Pipe RA & Pipe Dec & Offset & BAT $T_{90}$ & Pipe Dur. & Pipe SNR \\
 & (UTC) & (deg) & (deg) & (deg) & (deg) & (deg) & (s) & (s) & ($\sigma$) \\
\hline
\endfirsthead
\caption[]{Sample of Jointly Detected GRBs (continued).} \\
\hline
GRB Name & Pipeline Trigger Time & BAT RA & BAT Dec & Pipe RA & Pipe Dec & Offset & BAT $T_{90}$ & Pipe Dur. & Pipe SNR \\
 & (UTC) & (deg) & (deg) & (deg) & (deg) & (deg) & (s) & (s) & ($\sigma$) \\
\hline
\endhead
\hline
\endfoot
GRB130515A & 2013-05-15 01:21:17.909 & 283.44 & -54.28 & 290.23 & -50.04 & 5.94 & 0.30 & 0.133 & 42.8 \\
GRB130612A & 2013-06-12 03:22:22.997 & 259.77 & 16.72 & 259.02 & 20.97 & 4.31 & 4.00 & 2.671 & 16.8 \\
GRB130626A & 2013-06-26 10:51:03.839 & 273.12 & -9.52 & 265.37 & 0.18 & 12.40 & 0.16 & 0.179 & 13.4 \\
GRB130912A & 2013-09-12 08:34:58.562 & 47.61 & 14.00 & 44.31 & 9.06 & 5.90 & 0.28 & 1.466 & 24.5 \\
GRB131004A & 2013-10-04 21:41:04.100 & 296.11 & -2.95 & 298.85 & -7.45 & 5.26 & 1.54 & 1.466 & 21.8 \\
GRB131031A & 2013-10-31 11:33:34.658 & 29.62 & -1.62 & 34.69 & -5.54 & 6.40 & 6.40 & 2.671 & 38.4 \\
GRB131128A & 2013-11-28 15:06:25.263 & 355.31 & 31.29 & 2.75 & 27.10 & 7.74 & 3.00 & 1.979 & 19.5 \\
GRB140320A & 2014-03-20 02:12:46.025 & 281.84 & -11.19 & 279.73 & -5.70 & 5.88 & 0.51 & 0.327 & 17.3 \\
GRB140402A & 2014-04-02 00:10:06.975 & 207.63 & 5.99 & 200.90 & 7.90 & 6.95 & 0.90 & 0.098 & 18.7 \\
GRB140408A & 2014-04-08 13:15:56.655 & 290.72 & -12.58 & 286.55 & -14.23 & 4.38 & 3.63 & 6.573 & 15.0 \\
GRB140606A & 2014-06-06 10:58:13.562 & 201.80 & 37.60 & 222.75 & 37.48 & 16.58 & 0.34 & 0.327 & 13.0 \\
GRB140713A & 2014-07-13 18:43:46.390 & 281.14 & 59.62 & 288.46 & 60.51 & 3.76 & 6.02 & 2.671 & 24.2 \\
GRB140901B & 2014-09-01 06:17:25.588 & 112.18 & -29.21 & 109.30 & -31.05 & 3.10 & 6.40 & 6.573 & 17.6 \\
GRB141004A & 2014-10-04 23:20:54.606 & 76.72 & 12.83 & 64.79 & 11.62 & 11.72 & 3.92 & 0.441 & 32.7 \\
GRB141005A & 2014-10-05 05:13:08.095 & 291.11 & 36.10 & 301.28 & 41.89 & 9.79 & 3.47 & 2.671 & 52.6 \\
GRB141205A & 2014-12-05 08:05:17.734 & 92.86 & 37.88 & 95.05 & 26.64 & 11.38 & 1.66 & 0.596 & 20.0 \\
GRB141229A & 2014-12-29 11:49:00.370 & 72.45 & -19.25 & 79.39 & -16.37 & 7.21 & 6.40 & 1.086 & 55.7 \\
GRB150101A & 2015-01-01 06:28:53.735 & 312.59 & 36.67 & 309.79 & 36.77 & 2.24 & 0.06 & 0.022 & 11.0 \\
GRB150101B & 2015-01-01 15:23:34.467 & 188.04 & -10.98 & 183.92 & -8.13 & 4.97 & 0.01 & 0.030 & 20.2 \\
GRB150120A & 2015-01-20 02:57:46.874 & 10.33 & 33.98 & 358.28 & 26.82 & 12.61 & 1.20 & 1.086 & 14.6 \\
GRB150301A & 2015-03-01 01:04:28.645 & 244.28 & -48.73 & 223.92 & -33.53 & 21.46 & 0.48 & 0.016 & 22.9 \\
GRB150831B & 2015-08-31 22:19:30.027 & 271.03 & -27.24 & 280.54 & -28.93 & 8.55 & 6.18 & 6.573 & 47.9 \\
GRB151031A & 2015-10-31 05:50:32.949 & 83.19 & -39.12 & 82.90 & -34.46 & 4.67 & 5.00 & 4.869 & 19.3 \\
GRB151228A & 2015-12-28 03:05:12.512 & 214.02 & -17.67 & 216.24 & -18.10 & 2.16 & 0.28 & 0.242 & 23.8 \\
GRB151229A & 2015-12-29 06:50:28.451 & 329.36 & -20.73 & 329.24 & -21.39 & 0.67 & 1.44 & 1.466 & 56.4 \\
GRB160408A & 2016-04-08 06:25:43.963 & 122.56 & 71.13 & 115.88 & 69.01 & 3.11 & 0.32 & 0.327 & 49.7 \\
GRB160411A & 2016-04-11 01:28:52.137 & 349.37 & -40.26 & 344.74 & -30.20 & 10.75 & 0.36 & 1.466 & 10.9 \\
GRB160424A & 2016-04-24 11:49:08.710 & 319.48 & -60.41 & 321.47 & -54.86 & 5.65 & 6.46 & 6.573 & 70.0 \\
GRB160612A & 2016-06-12 20:12:47.604 & 348.36 & -25.37 & 353.15 & -32.40 & 8.18 & 0.25 & 0.179 & 53.5 \\
GRB160624A & 2016-06-24 11:27:01.406 & 330.21 & 29.66 & 337.34 & 20.45 & 11.24 & 0.19 & 0.242 & 25.9 \\
GRB160714A & 2016-07-14 02:19:15.706 & 234.48 & 63.81 & 231.57 & 55.83 & 8.11 & 0.35 & 0.327 & 13.9 \\
GRB160726A & 2016-07-26 01:34:08.338 & 98.81 & -6.62 & 91.45 & -7.03 & 7.32 & 0.73 & 0.179 & 51.5 \\
GRB160821B & 2016-08-21 22:29:13.348 & 280.00 & 62.39 & 277.51 & 58.48 & 4.09 & 0.48 & 0.098 & 17.0 \\
GRB161001A & 2016-10-01 01:05:17.753 & 71.92 & -57.26 & 88.72 & -58.16 & 9.00 & 2.60 & 2.671 & 49.5 \\
GRB170127B & 2017-01-27 15:13:28.784 & 20.00 & -30.34 & 20.10 & -31.40 & 1.06 & 0.51 & 0.098 & 21.2 \\
GRB170318B & 2017-03-18 15:27:52.858 & 284.29 & 6.32 & 291.22 & 18.01 & 13.51 & 1.07 & 1.086 & 11.9 \\
GRB170325A & 2017-03-25 07:56:58.081 & 127.48 & 20.53 & 124.45 & 17.19 & 4.40 & 0.33 & 0.179 & 23.1 \\
GRB170803A & 2017-08-03 17:30:27.709 & 174.95 & -16.31 & 175.49 & -17.79 & 1.57 & 3.82 & 1.466 & 47.8 \\
GRB171103A & 2017-11-03 23:10:32.164 & 249.51 & -10.21 & 257.10 & -9.71 & 7.49 & 4.27 & 0.242 & 100.0 \\
GRB180402A & 2018-04-02 09:44:59.401 & 251.92 & -14.93 & 204.97 & 41.36 & 71.05 & 0.18 & 0.098 & 46.8 \\
GRB180418A & 2018-04-18 06:44:06.463 & 170.13 & 24.93 & 160.85 & 26.17 & 8.46 & 4.41 & 0.327 & 15.0 \\
GRB180715A & 2018-07-15 18:07:06.563 & 235.09 & -0.90 & 230.43 & -2.17 & 4.83 & 0.68 & 3.606 & 19.1 \\
GRB180718A & 2018-07-18 01:57:44.601 & 336.02 & 2.79 & 331.67 & 23.85 & 21.47 & 0.08 & 0.133 & 18.0 \\
GRB180727A & 2018-07-27 14:15:29.155 & 346.64 & -63.07 & 358.03 & -59.13 & 6.75 & 1.06 & 1.086 & 32.3 \\
GRB181126A & 2018-11-26 09:54:09.982 & 152.35 & -29.70 & 149.29 & -29.31 & 2.69 & 2.09 & 0.327 & 20.0 \\
GRB181202A & 2018-12-02 06:36:27.870 & 280.74 & 27.98 & 281.76 & 25.97 & 2.20 & 6.56 & 6.573 & 9.6 \\
GRB190109B & 2019-01-09 11:56:09.663 & 55.55 & 63.60 & 70.79 & 58.50 & 8.93 & 6.51 & 1.086 & 11.4 \\
GRB190326A & 2019-03-26 07:35:28.921 & 341.65 & 39.91 & 351.76 & 47.75 & 10.69 & 0.08 & 0.022 & 40.6 \\
GRB190331A & 2019-03-31 02:14:38.206 & 28.57 & 27.64 & 40.30 & 26.67 & 10.47 & 4.34 & 1.466 & 33.3 \\
GRB190427A & 2019-04-27 04:34:15.080 & 280.22 & 40.31 & 264.80 & 43.24 & 11.85 & 0.19 & 0.040 & 39.1 \\
GRB190627A & 2019-06-27 11:18:32.145 & 244.87 & -5.29 & 235.02 & -9.21 & 10.53 & 2.69 & 1.086 & 11.0 \\
GRB200411A & 2020-04-11 04:29:02.566 & 47.69 & -52.32 & 64.18 & -54.89 & 10.09 & 0.22 & 0.179 & 22.6 \\
GRB200416A & 2020-04-16 07:05:19.253 & 335.69 & -7.53 & 332.90 & -2.42 & 5.82 & 6.05 & 6.573 & 35.2 \\
GRB200809B & 2020-08-09 15:41:27.428 & 15.89 & -73.83 & 15.84 & -68.41 & 5.42 & 4.20 & 0.804 & 27.8 \\
GRB201006A & 2020-10-06 01:17:52.445 & 61.91 & 65.15 & 52.93 & 73.30 & 8.73 & 0.49 & 0.804 & 20.2 \\
GRB201221D & 2020-12-21 23:06:34.359 & 171.05 & 42.15 & 161.35 & 50.95 & 11.02 & 0.16 & 0.022 & 16.2 \\
GRB210119A & 2021-01-19 02:54:09.832 & 282.82 & -61.77 & 287.53 & -63.52 & 2.78 & 0.06 & 0.054 & 25.1 \\
GRB210217A & 2021-02-17 23:25:42.570 & 97.56 & 68.71 & 129.97 & 68.40 & 11.72 & 4.00 & 1.086 & 12.5 \\
GRB210308A & 2021-03-08 06:37:59.969 & 67.08 & 37.43 & 60.03 & 39.02 & 5.76 & 6.19 & 3.606 & 152.4 \\
GRB210730A & 2021-07-30 04:57:30.474 & 149.57 & 69.70 & 146.76 & 69.97 & 1.00 & 3.86 & 2.671 & 67.7 \\
GRB220412B & 2022-04-12 17:06:49.032 & 320.76 & -0.26 & 353.93 & 2.56 & 33.29 & 0.14 & 0.022 & 38.9 \\
GRB221120A & 2022-11-20 21:29:28.119 & 41.36 & 43.24 & 39.56 & 40.66 & 2.91 & 0.79 & 0.596 & 23.3 \\
GRB221226B & 2022-12-26 22:41:21.438 & 22.90 & -41.51 & 18.51 & -46.85 & 6.19 & 3.44 & 2.671 & 32.9 \\
GRB230228A & 2023-02-28 05:50:50.972 & 18.40 & 44.50 & 21.75 & 50.23 & 6.17 & 2.18 & 1.086 & 15.7 \\
GRB230723B & 2023-07-23 11:42:35.780 & 250.39 & -5.32 & 249.51 & -8.44 & 3.24 & 5.76 & 6.573 & 41.0 \\
GRB230903A & 2023-09-03 17:22:57.097 & 9.90 & -40.91 & 353.48 & -23.94 & 21.83 & 2.54 & 1.086 & 9.9 \\
GRB240109A & 2024-01-09 04:01:35.192 & 301.32 & 16.54 & 312.66 & 18.63 & 11.00 & 0.08 & 0.327 & 12.7 \\
GRB241115A & 2024-11-15 13:18:26.306 & 86.77 & -0.67 & 84.96 & 6.39 & 7.28 & 3.70 & 3.606 & 61.4 \\
GRB241209A & 2024-12-09 05:35:44.179 & 155.37 & 6.35 & 155.33 & 9.02 & 2.67 & 0.26 & 0.242 & 12.0 \\
GRB241229B & 2024-12-29 01:36:17.517 & 274.14 & 74.69 & 277.39 & 74.69 & 0.86 & 5.05 & 3.606 & 50.9 \\
GRB250128B & 2025-01-28 16:22:53.766 & 231.42 & -0.55 & 250.60 & 20.85 & 28.45 & 0.48 & 0.098 & 10.4 \\
GRB250321A & 2025-03-21 00:42:40.641 & 295.10 & 21.05 & 298.93 & 25.20 & 5.44 & 6.21 & 3.606 & 31.6 \\
\hline
\multicolumn{10}{p{0.95\textwidth}}{\footnotesize \textit{Note.} This table lists the parameters for the subset of GRBs shown in Figure 12. It highlights the localization offset between the Swift/BAT reference coordinates and our pipeline's maximum likelihood estimates, alongside a comparison of the reported durations.} \\
\end{longtable}
\bsp	
\label{lastpage}
\end{document}